\documentclass[5p]{elsarticle}

\usepackage{mathtools}
\DeclarePairedDelimiter\ceil{\lceil}{\rceil}

\usepackage{lipsum}

\usepackage{flushend}
\usepackage{tikz}
\usepackage{tikz-qtree}
\usetikzlibrary{trees}
\usepackage{soul}
\usepackage{lineno}
\usepackage[figurename=Figure]{caption}
\usepackage[tablename=Table]{caption}
\usepackage{amsfonts}
\usepackage{commath}
\usepackage{algorithm}
\usepackage{stfloats}
\usepackage{algpseudocode}
\usepackage{multicol}
\usepackage{graphicx}
\usepackage{subcaption}
\usepackage{mathtools}
\usepackage{filecontents}
\usepackage{balance}
\usepackage{multirow}
\usepackage{xcolor}
\usepackage{rotating}
\modulolinenumbers[5]
\journal{Journal of XXX}

\begin{document}
	
	\begin{frontmatter}
		
		\title{Resource-sharing Policy in Multi-tenant \\Scientific Workflow-as-a-Service Cloud Platform}
		
		\author[]{Muhammad H. Hilman\corref{mycorrespondingauthor}}
		\cortext[mycorrespondingauthor]{Corresponding author}
		\ead{hilmanm@student.unimelb.edu.au}
		
		\author[]{Maria A. Rodriguez}
		
		\author[]{Rajkumar Buyya}
		
		\address{\normalsize{Cloud Computing and Distributed Systems (CLOUDS) Laboratory \\School of Computing and Information Systems \\The University of Melbourne, Australia}}
		
		\begin{abstract}
			Increased adoption of scientific workflows in the community has urged for the development of multi-tenant platforms that provide these workflow executions as a service. As a result, Workflow-as-a-Service (WaaS) concept has been created by researchers to address the future design of Workflow Management Systems (WMS) that can serve a large number of users from a single point of service. These platforms differ from traditional WMS in that they handle a workload of workflows at runtime. A traditional WMS is usually designed to execute a single workflow in a dedicated process while WaaS cloud platforms enhance the process by exploiting multiple workflows execution in a multi-tenant environment model. In this paper, we explore a novel resource-sharing policy to improve system utilization and to fulfill various Quality of Service (QoS) requirements from multiple users in WaaS cloud platforms. We propose an \textbf{E}lastic \textbf{B}udget-constrained resource \textbf{P}rovisioning and \textbf{S}cheduling algorithm for \textbf{M}ultiple workflows that can reduce the computational overhead by encouraging resource sharing to minimize workflows' makespan while meeting a user-defined budget. Our experiments show that the EBPSM algorithm can utilize the resource-sharing policy to achieve higher performance in terms of minimizing the makespan compared to the state-of-the-art budget-constraint scheduling algorithm.
		\end{abstract}
		
	\end{frontmatter}
	
	
\section{Introduction}

Scientific workflows have accelerated the triumph of scientific missions on today's multi-discipline sciences \cite{doi:10.1177/1094342017704893}. This technology orchestrates and automates scientific applications in a way that reduces the complexity of managing scientific experiments. These workflows are composed of numerous tasks that are interconnected by data or control dependencies. Furthermore, scientific workflows are widely known by their requirements of extensive computational resources. Therefore, these resource-intensive applications are deployed in distributed systems with a high capacity of storage, network, and computing power to achieve a reasonable processing time.

Cloud computing has become a beneficial infrastructure for deploying scientific workflows. Cloud services, especially Infrastructure as a Service (IaaS) offerings, provide a pseudo-infinite pool of resources that can be leased on-demand with a pay-per-use scheme. This pay-as-you-go model substantially eliminates the need for having an upfront investment for massive computational resources. IaaS provides virtualized computational resources in the form of Virtual Machines (VM) with pre-defined CPU, memory, storage, and bandwidth in various pre-configured bundles (i.e., VM types). The users can then elastically acquire and release as many VMs as they need, and the providers generally charge the resource usage per time slot (i.e., billing period).

Designing algorithms for scheduling scientific workflow executions in clouds is not trivial. The clouds natural features evoke many challenges involving the strategy to decide what type of VMs should be provisioned and when to acquire and release the VMs to get the most efficient scheduling result. The trade-off between having a faster execution time and an economical cost is something that must be carefully considered in leasing a particular cloud instance \cite{Deelman:2008:CDS:1413370.1413421}. Other challenging factors are performance variation of VMs and uncertain overhead delays that might arise from the virtualized backbone technology of clouds, geographical distribution, and multi-tenancy \cite{Leitner:2016:PCS:2926746.2885497}.

These problems have attracted many computer scientists into cloud workflow scheduling research to fully utilize the clouds' capabilities for efficient scientific workflows execution \cite{ALKHANAK20161} \cite{CPE:CPE4041}. The majority of those studies focus on the scheduling of a single workflow in cloud environments. In this model, they assume a single user utilizes a Workflow Management System (WMS) to execute a particular workflow's job in the cloud. The WMS manages the execution of the workflow so that it can be completed within the defined Quality of Service (QoS) requirements. Along with the growing trend of scientific workflows adoption in the community, there is a need for platforms that provide scientific workflows execution as a service.

\begin{figure}[!t]
	\centering
	\begin{subfigure}[b]{0.215\textwidth}
		\includegraphics[width=\textwidth]{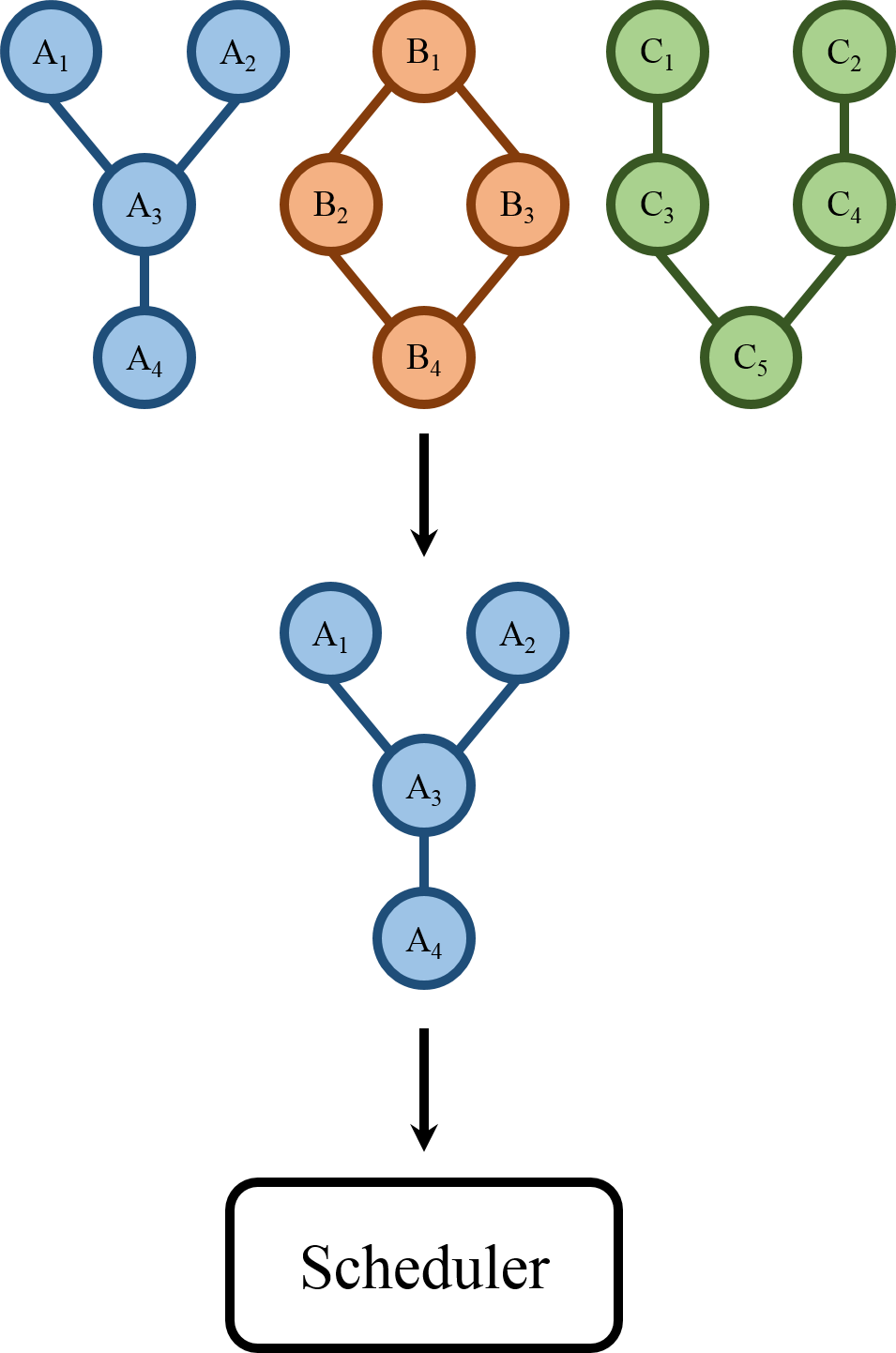}
		\subcaption{Independent}    
		\label{fig:indepenttask}
	\end{subfigure}    
	\quad
	\begin{subfigure}[b]{0.21\textwidth}
		\includegraphics[width=\textwidth]{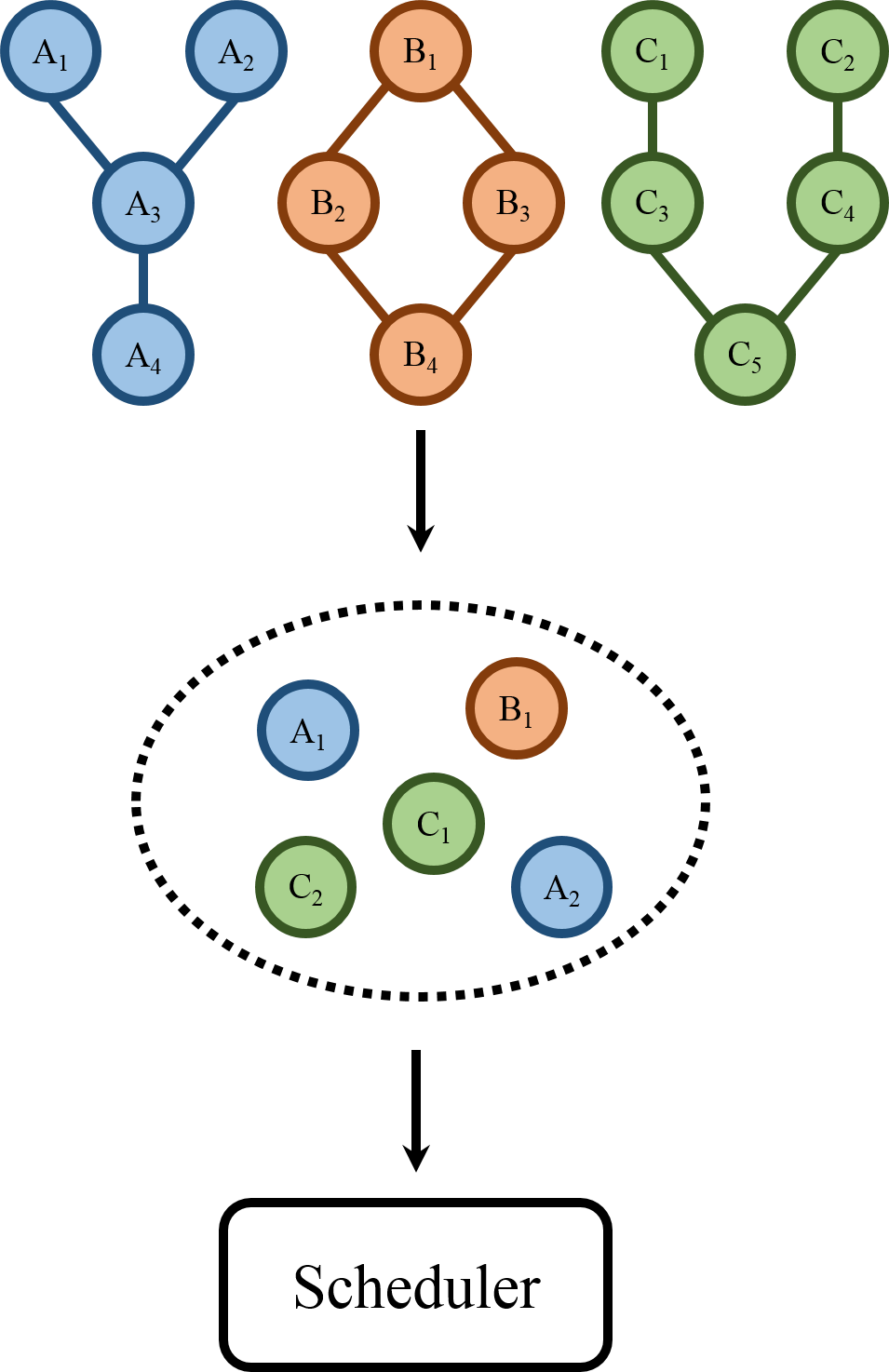}
		\subcaption{Simultaneous}
		\label{fig:intertask}
	\end{subfigure}
	\vspace{-1mm}
	\caption{Two approaches on scheduling multiple workflows}
	\vspace{-3mm}
\end{figure}

Workflow-as-a-Service (WaaS) is an evolving idea that offers workflows execution as a service to the scientific community. The platforms extend WMS technology that is commonly used for handling an individual execution of scientific workflow to serve a more significant number of users through a single point of service. Although some of the traditional cloud workflow scheduling algorithms can be extended for this problem, the specific nature of WaaS environments creates a set of particular needs that should be addressed. As an example, WaaS cloud platforms continuously receive many workflows' jobs from different users with their unique QoS requirements. The providers must be able to process these requests in a way that each of the requirements is fulfilled. A traditional WMS may process the workflows individually in a dedicated set of VMs as depicted in Figure \ref{fig:indepenttask}. This approach, after all, is the simplest way to ensure the QoS fulfillment of each job.

In this dedicated service scenario, a WMS manages different types of tasks' execution by tailoring their specific software configurations and requirements to a VM image. The VM containing this image then can be quickly deployed whenever a particular workflow is submitted. However, this model cannot easily be implemented in WaaS cloud platforms where many users with different workflow applications are involved. We cannot naively simplify the assumption where every VM image can be shared between multiple users with different requirements. Multiple workflow applications may need different software configurations, which implies a possible dependency conflict if they are fitted within a VM image. This assumption also creates a more complex situation where, at any given time, a new workflow application type needs to be deployed. This newly submitted job cannot reuse the already provisioned VMs as they may not contain its software configurations. Furthermore, dedicating a single workflow execution in a set of VMs is considered inefficient as it leads to the inevitable schedule gaps from inter-dependent workflow tasks that result in the VMs being underutilized.

Therefore, adopting an appropriate resource-sharing policy and at the same time scheduling multiple workflows simultaneously, as shown in Figure \ref{fig:intertask}, is considerably preferred for multi-tenant WaaS cloud platforms. We argue that introducing this strategy creates a more efficient multi-tenant platform as it reduces the unnecessary overhead during the execution. The efficiency may be gained from sharing the same workflow application software for different users by tailoring a specific configuration in a container image instead of a VM image. This strategy enables (i) sharing and reusing the already provisioned VMs between users to utilize the inevitable scheduling gaps from intra-workflow's dependency and (ii) sharing local cached images and datasets within a VM that creates a locality, which eliminates the need for network-related activities before the execution.

Based on these described problems and requirements, we propose EBPSM, an \textbf{E}lastic \textbf{B}udget-constrained resource \textbf{P}rovisioning and \textbf{S}cheduling algorithm for \textbf{M}ultiple workflows designed for WaaS cloud platforms. A further elaboration of the resource sharing policy that has been introduced in \cite{RODRIGUEZ2018739} by utilizing container technology to share computational resources (i.e., VMs). This new proposed algorithm focuses more on a budget-constrained scheduling scenario where its budget distribution strategy has been discussed in \cite{8109130}.

The EBPSM algorithm can make a quick decision to schedule the workflow tasks dynamically and empower the sharing of software configuration and reuse of already provisioned VMs between users using container technology. Our algorithm also considers inherent features of clouds that affect multiple workflows scheduling, such as performance variation of VMs \cite{Leitner:2016:PCS:2926746.2885497} and provisioning delay \cite{7761629} into the VMs auto-scaling policy. Furthermore, EBPSM implements an efficient budget distribution strategy that allows the algorithm to provision the fastest VMs possible to minimize the makespan and adopt the container images and datasets sharing policy to eliminate the need for network transfer between tasks' execution. Our extensive experiments show that EBPSM which adopts the resource-sharing policy can significantly reduce the overhead, which implies the minimization of workflows' makespan.

The rest of this paper is organized as follows: Section \ref{section:relatedwork} reviews works that are related to this proposal. Section \ref{section:problemform} explains the problem formulation of multiple workflow scheduling in WaaS cloud platforms, including the assumption of application and resource models. The proposed algorithm is described in Section \ref{section:algorithm} followed by the performance evaluation in Section \ref{section:performanceeval}. Finally, the conclusion and future work are elaborated in Section \ref{section:conclusion}.

\section{Related Work}\label{section:relatedwork}

Scheduling jobs in distributed systems is an extensive research area \cite{7425222}. In general, the solution for specific jobs that comes from multiple users/sources utilized the online schedulers which implemented the lightweight approaches such as combinatorial search \cite{4180346} and heuristics algorithm \cite{10.1145/1998582.1998637}. The similar online approaches are also used for managing the workflows jobs, which avoid the high computational complexity from the overhead and performance variation in dynamic environments. In this case, the problem of scheduling multiple workflows is facing both the challenges from the unique structure of the jobs and the opportunities to further utilized the systems which involves the resource-sharing of multiple users.

The majority of works in multiple workflows scheduling have pointed out the necessity of reusing already provisioned VMs to reduce the idle gaps and increase system utilization. Examples include the CWSA \cite{7457258} that uses a depth-first search technique to find potential schedule gaps between tasks' execution. Another work is the CERSA \cite{Chen2018} that dynamically adjusts the VM allocation for tasks in a reactive fashion whenever a new workflow job is submitted to the system. These works' idea to fill the schedule gaps between tasks' execution of a workflow to be utilized for scheduling tasks from another workflow is similar to our proposal. However, they merely assume that different workflow applications could be deployed into any existing VM available without considering the possible complexity of software dependency conflicts. Our work differs in the way that we model the software configurations into a container image before deploying it to the VMs for execution.

The use of the container for deploying scientific workflows has been intensively researched. Examples include the work by Qasha et al. \cite{7830693} that deployed a TOSCA-based workflow\footnote{https://github.com/ditrit/workflows} using Docker\footnote{https://www.docker.com/} container on e-Science Central platform\footnote{https://www.esciencecentral.org/}. Although their work is done on a single VM, the result shows a promising future scientific workflows reproducibility using container technology. A similar result is presented by Liu et al. \cite{7600178} that convinces less overhead performance and high flexibility of deploying scientific workflows using Docker containers. Finally, the adCFS \cite{8171362} is designed to schedule containerized scientific workflows that encourage a CPU-sharing policy using a Markov-chain model to assign the appropriate CPU weight to containers. Those solutions are the early development of containerized scientific workflows execution. Their results show high feasibility to utilize container technology for efficiently bundling software configurations for workflows that are being proposed for WaaS cloud platforms.

One of the challenges of executing scientific workflows in the clouds is related to the data locality. The communication overhead for transferring the data between tasks' execution may take a considerable amount of time that might impact the overall makespan. A work by Stavrinides and Karatza \cite{STAVRINIDES2017120} shows that the use of distributed in-memory storage to store the datasets locally for tasks' execution can reduce the communication overhead. Our work is similar regarding the data locality policy to minimize the data transfer. However, we propose the use of cached datasets in VMs local storage to endorse the locality of data. We enhance this policy so that the algorithm can intelligently decide which task to be scheduled in specific VMs that can provide the minimum execution time given the available cached datasets.

Two conflicting QoS requirements in scheduling (e.g., time and cost) have been a significant concern when deploying scientific workflows in clouds. A more relaxed constraint to minimize the trade-off between these two requirements is shown in several works that consider scheduling workflows within the deadline and budget constraints. They do not attempt to optimize one or both of the QoS requirements but instead maximizing the success rate of workflows execution within the constraints. Examples of these works include MW-DBS \cite{ARABNEJAD2017120} and MW-HBDCS \cite{Zhou2018} algorithms. Another similar work is the MQ-PAS \cite{ARABNEJAD2017211} algorithm that emphasizes on increasing the providers' profit by exploiting the budget constraint as long as the deadline is not violated. Our work considers the user-defined budget constraint in the scheduling, but it differs in the way that the algorithm aims to optimize the overall makespan of workflows while meeting their budget.

Several works specifically focus on handling the real-time workload of workflows in WaaS cloud platforms. This type of workload raises uncertainty issue as the platforms have no knowledge of the arriving workflows. EDPRS \cite{Chen2017} adopts a dynamic scheduling approach using event-driven and periodic rolling strategies to handle the uncertainties in real-time workloads. Another work, called ROSA \cite{8443134}, controls the queued jobs--which increase the uncertainties along with the performance variation of cloud resources--in the WaaS cloud platforms to reduce the waiting time that can prohibit the uncertainties propagation. Both algorithms are designed to schedule multiple workflows dynamically to minimize the operational cost while meeting the deadline. Our algorithm has the same purpose of handling the real-time workload and reducing the effect of uncertainties in WaaS environments. However, we differ in that our scheduling objectives are minimizing the workflows' makespan while meeting the user-defined budget.

The majority of works in workflow scheduling that aim to minimize the makespan, while meeting the budget constraints, adopt a static scheduling approach. This approach finds a near-optimal solution of mapping the tasks to VMs--with various VM types configuration--to get a schedule plan before runtime. Examples of these works include MinMRW-MC \cite{8411028}, HEFT-Budg, and MinMin-Budg \cite{8425321}. However, this static approach is considered inefficient for WaaS cloud platforms as it increases the waiting time of arriving workflows due to the intensive pre-processing computation time to generate a schedule plan.

On the other hand, several works consider scheduling budget-constrained workflows dynamically driven by the available user-defined budget. In this area, examples include BAT \cite{7828507} which distributes the budget of a particular workflow to its tasks by trickling down the available budget based on the depth of tasks. Another work, MSLBL \cite{CHEN20171} algorithm distributes the budget by calculating a proportion of the sub-budget efficiently to reduce the unused budget. However, those solutions are designed for a single cloud workflow scheduling scenario. To the best of our knowledge, none of these type of budget-constrained algorithms, that aim to tackle the problem in multiple workflows--which resembles the problem in WaaS cloud platforms--has been proposed.

\section{Problem Formulation}\label{section:problemform}

\begin{figure}[!t]
	\centering \includegraphics[width=.925\linewidth]{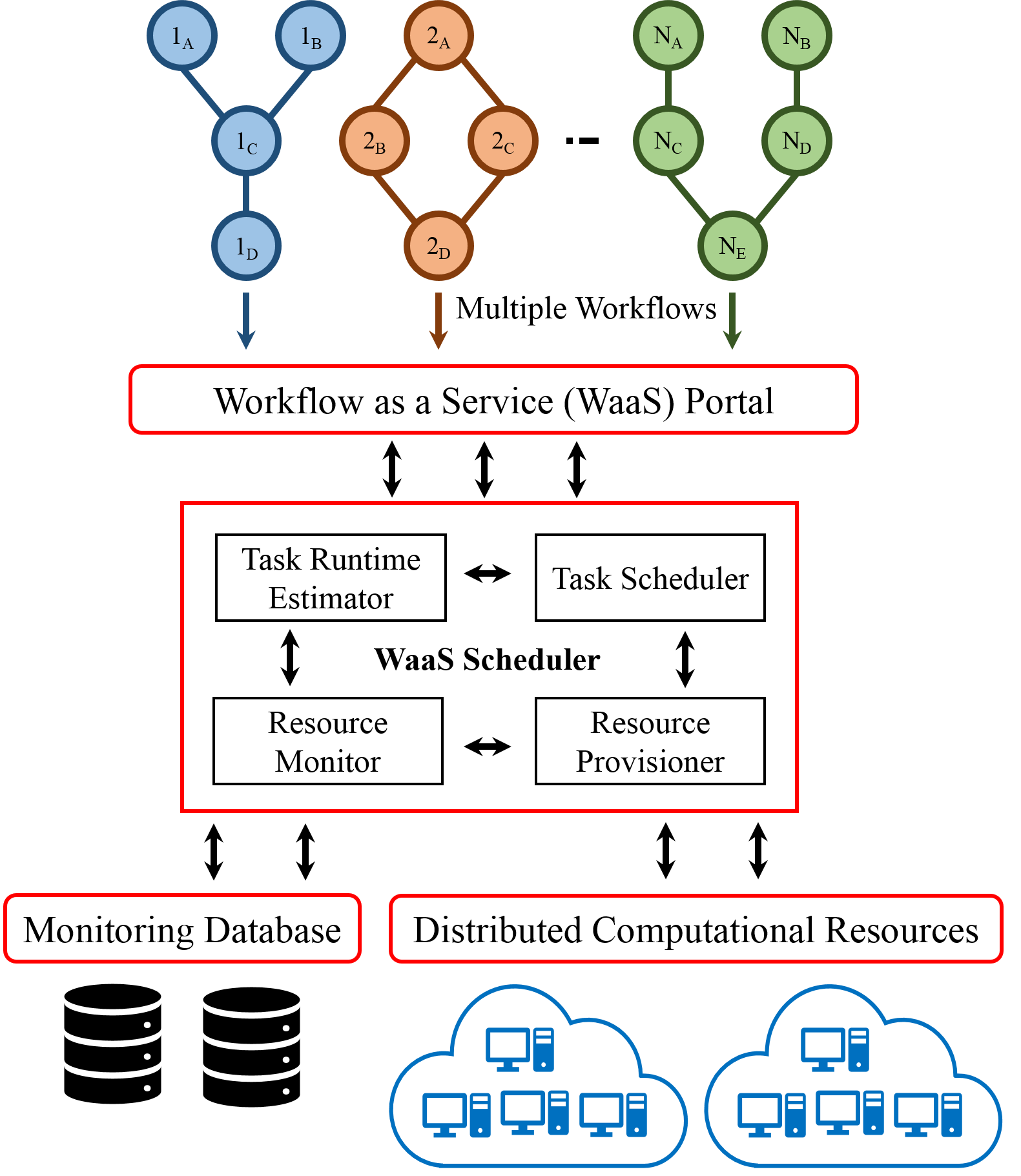}
	\caption{Architecture of Workflow-as-a-Service cloud platform}
	\label{figure:archit}
	\vspace{-4mm}
\end{figure}

In this paper, we consider the resource provisioning and scheduling algorithm to process multiple workflows simultaneously. The workflows that are submitted for execution may have the same workflow application type and require the same datasets, but they are not necessarily related to each other. Specifically, the workflows might belong to different users; therefore, they may have different QoS requirements. We assume that these workflows are being executed in a WaaS cloud platform, and a proposed architecture for this particular system is depicted in Fig: \ref{figure:archit}.

\subsection{Workflow-as-a-Service (WaaS) Cloud Platform}\label{subsection:waasplatforms}

WaaS cloud platforms can be placed either in the Platform as a Service (PaaS) or Software as a Service (SaaS) layer of the cloud service model. WaaS providers utilize distributed computational resources from Infrastructure as a Service (IaaS) services to serve the enormous need for computing power in the execution of scientific workflows. They provide an end-to-end service to scientists from the submission portal that allows users to define their requirements, automate their applications' installation and configuration based on the available or newly created templates, and stage input/output data from the system. Meanwhile, the rest of the data and resource management, including workflow scheduling and resource provisioning, is transparent to the users. This platform is designed to process multiple workflows from different users as they arrive continuously. Therefore, WaaS cloud platforms must handle a high level of complexity derived from their multi-tenant nature, in contrast to a traditional WMS that is generally deployed to manage a single workflow execution. In this work, we focus on the resource provisioner and task scheduler aspects that are designed to deal with the dynamic workload of workflows arriving into WaaS cloud platforms.

\subsection{Application and Resource Model}\label{subsection:appandresmodel}

We consider a workload of workflows that are modelled as DAGs (Directed Acyclic Graphs). Furthermore, a workload $W=\{w_1, w_2, w_3, . . ., w_n\}$ is composed of multiple workflows, where a workflow $w$ consists of a number of tasks $T=\{t_1, t_2, t_3, . . ., t_n\}$ and edges $E$. An edge $e_{ij}(t_i,t_j)$ represents the data dependency between task $t_i$ and $t_j$ where $t_j$, as a successor (i.e., child task), will only start the execution after $t_i$, as predecessor (i.e., parent task), is completed and output data $d_{t_i}^{out}$ from $t_i$ is available on a VM allocated for $t_j$ as input data $d_{t_j}^{in}$. The workflow is completed the execution when the exit task $t_{exit}$ (i.e., task without successors) is finished being processed. We also assume that each workflow $w$ is associated with a budget $\beta_w$ that is defined as a soft constraint of cost representing users' willingness to pay for the execution of the workflows.

The task $t$ is executed within a container--that bundles software configurations for a particular workflow in a container image--which is then deployed on VMs. A container provisioning delays $prov_c$ is acknowledged to download the image, setup, and initiate the container on an active VM. In this work, we set aside the idea of co-locating several running containers within a VM for further study, as the focus of the work is mainly to explore the sharing policy of VMs in terms of its computing, storage, and network capacities. Therefore, we assume that only one container can run on top of a VM at a particular time. Furthermore, the same host VM performance of CPU, memory, and bandwidth, can be achieved in a particular container. Once the container is deployed, VM local storage maintains its image so it can be reused without the need to re-download the containers. We assume WaaS scheduler sends custom signals by using commands in the container (e.g., Docker exec) to trigger tasks' execution within containers to avoid the necessity of container redeployment.

We consider a pay-as-you-go scheme in Iaas clouds, where VMs are provisioned on-demand and are priced per billing period $bp$ (i.e., per-second, per-minute, per-hour). Hence, any partial use of the VM is rounded up and charged based on the nearest $bp$. In this work, we assumed a fine-grained per-second $bp$ as it is lately being adopted by the majority of IaaS cloud providers including Amazon EC2\footnote{https://aws.amazon.com/about-aws/whats-new/2017/10/announcing-amazon-ec2-per-second-billing/}, Google Cloud\footnote{https://cloud.google.com/blog/products/gcp/extending-per-second-billing-in-google}, and Azure\footnote{https://azure.microsoft.com/en-us/pricing/details/virtual-machines/windows/}. We model a data center within a particular availability zone from a single IaaS cloud provider to reduce the network overhead and eliminate the cost associated with data transfer between zones. Our work considers a heterogeneous cloud environment model where VMs with different VM types $VMT = \{vmt_1, vmt_2, vmt_3, . . ., vmt_n\}$ which have various processing power $p_{vmt}$ and different cost per billing period $c_{vmt}$ can be leased. We consider that all types of VM always have an adequate memory capacity to execute the various type of workflows' tasks. Finally, we can safely assume that the VM type with a higher $p_{vmt}$ has more expensive $c_{vmt}$ than the less powerful and slower ones.

Each VM has a bandwidth capacity $b_{vmt}$ that is relatively the same between different VM types as they come from a single availability zone. We do not restrict the number of VMs to be provisioned during multiple workflows execution, but we also acknowledge the delay in acquiring VMs $prov_{vmt}$ from the IaaS provider. We assume that the VMs can be eliminated immediately from the WaaS cloud platform without additional delay. Furthermore, we consider performance variation of VMs that might come from a virtualized backbone technology of clouds, geographical distribution, and multi-tenancy \cite{Leitner:2016:PCS:2926746.2885497} and that CPU of VM advertised by IaaS provider is the highest CPU capacity that can be achieved by the VMs. We do not assume another performance degradation of using the containerized environments as Kozhirbayev and Sinnott \cite{KOZHIRBAYEV2017175} reported its near-native performance.

A global storage system $GS$ is modelled for data sharing between tasks (e.g., Amazon S3) with unlimited storage capacity. This global storage has reading rates $GS_r$ and writing rates $GS_w$ respectively. In this model, the tasks transfer their outputs from the VMs to the storage and retrieve their inputs from the same place before the execution. Therefore, the network overhead is inevitable and is one of the uncertainties in clouds as the network performance degradation can be observed due to the amount of traffic and the virtualized backbone \cite{6848061}. To reduce the need for accessing storage $GS$ for retrieving the data, VMs local storage $LS_{vmt}$ is also modelled to maintain $d_{t}^{in}$ and $d_{t}^{out}$ after particular tasks' execution using FIFO policy. It means the earliest stored data will be deleted whenever the capacity of $LS_{vmt}$ cannot accommodate new data needing to be cached. Furthermore, the time taken to retrieve the input data for a particular task's execution from global storage is shown in Eq. \ref{eq:datainput}. 
\begin{equation}
\label{eq:datainput}
T_{vmt}^{d_{t}^{in}}=d_{t}^{in}/GS_{r} + d_{t}^{in}/b_{vmt}
\end{equation}
It is worth noting that there is no need to transfer the input data from the global storage whenever it is available in the VM as cached data from previous tasks execution. Similarly, the time needed for storing the output data on the global storage is depicted in Eq. \ref{eq:dataoutput}.
\begin{equation}
\label{eq:dataoutput}
T_{vmt}^{d_{t}^{out}}= d_{t}^{out}/GS_{w} + d_{t}^{out}/b_{vmt}
\end{equation}

The runtime $RT_{vmt}^{t}$ of a task $t$ in a VM of type $vmt$ is assumed as available to the scheduler as part of the scheduling process. The fact is that this runtime can be estimated using various approaches including machine learning techniques \cite{8013738}\cite{8603156}, but we simplify the assumption where it is calculated based on the task's size $S_t$ in Million of Instruction (MI) and the processing capacity $p_{vmt}$ of the particular VM type in Million of Instruction Per Second (MIPS) as shown in Eq. \ref{eq:runtime}.
\begin{equation}
\label{eq:runtime}
RT^{t}_{vmt} = S_{t} / p_{vmt}
\end{equation}
It needs to be noted that this $RT^{t}_{vmt}$ value is only an estimate and the scheduler does not depend on it being 100\% accurate as it represents one of the uncertainties in clouds. Furthermore, a maximum processing time of a task in a VM type $PT_{vmt}^{t}$ consists of reading the input data required from the global storage, executing the task, and writing the output to the storage which are depicted in Eq. \ref{eq:makespan}.
\begin{equation}
\label{eq:makespan}
PT^{t}_{vmt} = T_{vmt}^{d_{t}^{in}} + RT^{t}_{vmt} + T_{vmt}^{d_{t}^{out}}
\end{equation}
From the previous equations, we can calculate the maximum cost $C_{vmt}^{t}$ of executing a task $t$ on a particular $vmt$ as shown in Eq. \ref{eq:cost}.
\begin{equation}
\label{eq:cost}
C_{vmt}^{t} =\ceil{(prov_{vmt} + prov_c + PT_{vmt}^{t}) / bp} * c_{vmt} 
\end{equation}

The budget-constrained scheduling problem that is being addressed in this paper is concerned with minimizing the makespan (i.e., actual finish time of $t_{exit}$) of each workflow while meeting the user-defined budget as depicted in Eq. \ref{eq:budget}.
\begin{equation}
\label{eq:budget}
min \quad AFT(t_{exit}) \quad \textrm{while} \quad \sum_{n=1}^{T} C_{vmt}^{t_n} \leq \beta_w 
\end{equation}
Intuitively, the budget $\beta_w$ will be spent efficiently on $PT_{vmt}^{t}$ if the overhead components $prov_{vmt}$ and $prov_c$ that burden the cost of a task's execution can be discarded. This implies to the minimization of $AFT(t_{exit})$ as the fastest VMs can be leased based on the available budget $\beta_w$. Another important note is that, further minimization can be achieved when the task $t$ is allocated to the VM with $d_{t}^{in}$ available, so the need for $T_{vmt}^{d_{t}^{in}}$ which is related to the network factor that becomes one of the sources of uncertainties can be eliminated.

However, it needs to be noted that there still exist some uncertainties in $RT^{t}_{vmt}$ as the estimate of $S_{t}$ is not entirely accurate, and the performance of VMs depicted in $PC_{vmt}$ can be degraded at any time. Hence, there must be a control mechanism to ensure that these uncertainties do not propagate throughout the tasks' execution and cause a violation of $\beta_w$. This control can be done by evaluating the real value of a task's execution cost $C_{vmt}^{t}$ right after the task is completed. In this way, its successors can adjust their sub-budget allocation so that the total cost will not violate the budget $\beta_w$.

\section{The EBPSM Algorithm}\label{section:algorithm}

\begin{algorithm}[!t]
	\small
	\caption{Budget Distribution}\label{algo:budgetdistribution}
	\begin{algorithmic}[1]
		\Procedure{distributeBudget(\textit{$\beta$, T})}{}
		\State{\textit{S} = tasks' estimated execution order}
		\For{each task \textit{t} $\in$ \textit{T}}
		\State{\textit{allocateLevel}(\textit{t, l})}
		\State{\textit{initiateBudget}(\textit{0, t})}
		\EndFor
		\For{each level \textit{l}}
		\State{\textit{$T_{l}$} = set of all tasks in level \textit{l}}
		\State{sort \textit{$T_{l}$} based on ascending Earliest Finish Time}
		\State{\textit{put}(\textit{$T_{l}$, S})}
		\EndFor
		\While{$\beta$ $>$ 0}
		\State{\textit{t} = \textit{S.poll}}
		\State{\textit{vmt} = chosen VM type}
		\State{\textit{allocateBudget}(\textit{C$^{t}_{vmt}$, t})}
		\State{$\beta$ $=$ $\beta$ $-$ \textit{C$^{t}_{vmt}$}}
		\EndWhile
		\EndProcedure
	\end{algorithmic}
\end{algorithm}

In this paper, we propose a dynamic heuristic resource provisioning and scheduling algorithm designed for WaaS cloud platforms to minimize the makespan while meeting the budget. The algorithm is developed to efficiently schedule scientific workflows in multi-tenant platforms that deal with dynamic workload heterogeneity, the potential of resource inefficiency, and uncertainties of overheads along with performance variations. Overall, the algorithm enhances the reuse of software configurations, computational resources, and datasets to reduce the overheads that become one of the critical uncertainties in cloud environments. This policy implements a resource-sharing model by utilizing container technology and VMs local storage in the decision-making process to schedule tasks and auto-scale the resources.

When a workflow is submitted to the WaaS portal, its owner may define the software requirements by creating a new container image or choosing an existing template. Whenever the user selects the existing images, the platform will identify the possibly relevant information that is maintained from the previous workflows' execution, including analyzing previous actual runtime execution and its related datasets. Furthermore, the user then defines the budget $\beta_w$ that is highly likely different from various users submitting the same type of workflow.

Next, the workflow is forwarded to the WaaS scheduler and is preprocessed to assign sub-budget for each task based on the user-defined $\beta_w$. This sub-budget along with the possible sharing of software configurations and datasets will lead the decisions made at runtime to schedule a task onto either an existing VM in the resource pool or a new VM provisioned from the IaaS cloud provider. The first phase of the budget distribution algorithm is to estimate the potential tasks' execution order within a workflow. The entry task(s) in the first level of a workflow are scheduled first, followed by their successors in the next level until it reaches the exit task(s). In this case, we assign every task to a level within a workflow's structure based on the Deadline Top Level (DTL) approach, as seen in Eq. \ref{eq:BTL}
\begin{equation}
\label{eq:BTL}
level(t) = 
\begin{cases}
0 & \textrm{if} \quad pred(t) = \emptyset\\
\max\limits_{p \in pred(t)}{level(p) + 1} & \textrm{otherwise}    \end{cases}
\end{equation}

\noindent
Furthermore, to determine the tasks' priority within a level, we sort them based on their Earliest Finish Time (EFT) in an ascending order as shown in Eq. \ref{eq:EFT}
\begin{equation}
\label{eq:EFT}
eft(t) = 
\begin{cases}
PT^{t}_{vmt} & \textrm{if} \quad pred(t) = \emptyset\\
\max\limits_{p \in pred(t)}{eft(p) + PT^{t}_{vmt}} & \textrm{otherwise}    \end{cases}
\end{equation}

\begin{algorithm}[!t]
	\small
	\caption{Scheduling}\label{algo:schedule}
	\begin{algorithmic}[1]
		\Procedure{scheduleQueuedTasks(\textit{q})}{}
		\State{sort \textit{q} by ascending Earliest Start Time (EST)}
		\While {\textit{q} is not empty}
		\State{\textit{t} $=$ \textit{q.poll}}
		\State{\textit{container} $=$ \textit{t.container}}
		\State{\textit{vm} $=$ \textit{null}}
		\If{there are idle VMs}
		\State{\textit{VM$_{idle}$} $=$ set of all idle VMs}
		\State{\textit{VM$_{idle}^{input}$} $=$ set of \textit{vm} $\in$ \textit{VM$_{idle}$} that have}
		\Statex{\hspace{\algorithmicindent}\hspace{\algorithmicindent}\hspace{\algorithmicindent}\hspace{\algorithmicindent}\textit{t}'s input data}
		\State{\textit{vm} $=$ \textit{vm} $\in$ \textit{VM$_{idle}^{input}$} that can finish \textit{t} within}
		\Statex{\hspace{\algorithmicindent}\hspace{\algorithmicindent}\hspace{\algorithmicindent}\hspace{\algorithmicindent}\textit{t.budget} with the fastest time}
		\If{\textit{vm} $=$ \textit{null}}
		\State{\textit{VM$_{idle}$} $=$ \textit{VM$_{idle}$} $\setminus$ \textit{VM$_{idle}^{input}$}}
		\State{\textit{VM$_{idle}^{container}$} $=$ set of \textit{vm} $\in$ \textit{VM$_{idle}$} that have}
		\Statex{\hspace{\algorithmicindent}\hspace{\algorithmicindent}\hspace{\algorithmicindent}\hspace{\algorithmicindent}\hspace{\algorithmicindent}\textit{container} deployed}
		\State{\textit{vm} $=$ \textit{vm} $\in$ \textit{VM$_{idle}^{container}$} that can finish \textit{t}}
		\Statex{\hspace{\algorithmicindent}\hspace{\algorithmicindent}\hspace{\algorithmicindent}\hspace{\algorithmicindent}\hspace{\algorithmicindent} within \textit{t.budget} with the fastest time}
		\If{\textit{vm} $=$ \textit{null}}
		\State{\textit{VM$_{idle}$} $=$ \textit{VM$_{idle}$} $\setminus$ \textit{VM$_{idle}^{container}$}}
		\State{\textit{vm} $=$ \textit{vm} $\in$ \textit{VM$_{idle}$} that can finish \textit{t}}
		\Statex{\hspace{\algorithmicindent}\hspace{\algorithmicindent}\hspace{\algorithmicindent}\hspace{\algorithmicindent}\hspace{\algorithmicindent}\hspace{\algorithmicindent} within \textit{t.budget} with the fastest time}
		\EndIf
		\EndIf
		\Else
		\State{\textit{vmt} $=$ fastest VM type within \textit{t.budget}}
		\State{\textit{vm} $=$ \textit{provisionVM}(\textit{vmt})}
		\EndIf
		\If{\textit{vm} $!=$ \textit{null}}
		\If{\textit{vm.container $!=$ container}}
		\State{\textit{deployContainer(vm, container)}}
		\EndIf
		\State{\textit{scheduleTask}(\textit{t, vm})}
		\EndIf
		\EndWhile
		\EndProcedure
	\end{algorithmic}
\end{algorithm}

After estimating the possible tasks' execution order, the algorithm iterates over the tasks based on this order and distributes the budget to each task. This budget distribution algorithm estimates the sub-budget of a task based on the cost $c_{vmt}$ of particular VM types. At first, the algorithm chooses VMs with the cheapest types for the task. Whenever there is any extra budget left after all tasks get their allocated sub-budgets, the algorithm uses this extra budget to upgrade the sub-budget allocation for a faster VM type starting from the earliest tasks in the order. This approach is called the Slowest First Task-based Budget Distribution (SFTD) \cite{8109130}. The detailed strategy is depicted in Algorithm \ref{algo:budgetdistribution}.

Once a workflow is preprocessed, and its budget is distributed, WaaS scheduler can begin the scheduling process, this step is illustrated in Algorithm \ref{algo:schedule}. The primary objective of this scheduling algorithm is to reuse the VMs that have datasets and containers--with software configurations available--in VMs local storage that may significantly reduce the overhead of retrieving the particular input data and container images from $GS$. This way, the algorithm avoids the provisioning of new VMs as much as possible, which reduces the VM provisioning delay and minimizes the network communication overhead from data transfer and downloading container images that contribute to the uncertainties in the WaaS environments.

Furthermore, WaaS scheduler releases all the entry tasks (i.e., tasks with no parents) of multiple workflows into the queue. As the tasks' execution proceeds, the child tasks--whose parents are completed--become ready for execution and are released into the scheduling queue. As a result, at any point in time, the queue contains all the tasks from different workflows submitted to the WaaS cloud platform that is ready for execution. The queue is periodically being updated whenever one of the two events triggered the scheduling cycle. Those are the arrival of a new workflow's job and the completion of a task's execution.

\begin{algorithm}[!t]
	\small
	\caption{Budget Update}\label{algo:budgetupdate}
	\begin{algorithmic}[1]
		\Procedure{updateBudget(\textit{T})}{}
		\State{$t_{f} =$ completed task}
		\State{\textit{T$_{u}$} = set of unscheduled \textit{t} $\in$ \textit{T}}
		\State{$\beta_{u}$ = total sum of \textit{t.budget}, where \textit{t} $\in$ \textit{T$_{u}$}}
		\State{\textit{sb} = spare budget}
		\If{\textit{C$^{t_{f}}_{vmt}$} $\leq$ (\textit{t$_{f}$.budget} $+$ \textit{sb})}
		\State{\textit{sb} = (\textit{t$_{f}$.budget} $+$ \textit{sb}) $-$ \textit{C$^{t_{f}}_{vmt}$}}
		\State{$\beta_{u}$ = $\beta_{u}$ + $sb$}
		\Else
		\State{\textit{debt} = \textit{C$^{t_{f}}_{vmt}$} $-$ (\textit{t$_{f}$.budget} $+$ \textit{sb})}
		\State{$\beta_{u}$ = $\beta_{u}$ $-$ \textit{debt}}
		\EndIf
		\State{\textsc{distributeBudget}(\textit{$\beta_{u}$, \textit{T$_{u}$}})}
		\EndProcedure
	\end{algorithmic}
\end{algorithm}

\begin{algorithm}[!t]
	\small
	\caption{Resource Provisioning}\label{algo:resprov}
	\begin{algorithmic}[1]
		\Procedure{manageResource}{}
		\State{\textit{VM$_{idle}$} $=$ all leased VMs that are currently idle}
		\State{\textit{threshold$_{idle}$} $=$ idle time threshold}
		\For{each \textit{vm$_{idle}$} $\in$ \textit{VM$_{idle}$}}
		\State{\textit{t$_{idle}$} $=$ idle time of \textit{vm}}
		\If{\textit{t$_{idle}$} $\geq$ \textit{threshold$_{idle}$}}
		\State{terminate \textit{vm$_{idle}$}}
		\EndIf
		\EndFor
		\EndProcedure
	\end{algorithmic}
\end{algorithm}

In every scheduling cycle, each task in the scheduling queue is processed as follows. The first step is to find a set of $VM_{idle}$ on the system that can finish the task's execution with the fastest time within its budget. The algorithm estimates the execution time by not only calculating $PT^{t}_{vmt}$ but also considering possible overhead $prov_c$ caused by the need for initiating a $container$ in case the available $VM_{idle}$ does not have a suitable $container$ deployed.

At first attempt, $VM_{idle}$ with input datasets $VM_{idle}^{input}$ are preferred. The $VM_{idle}^{input}$ that has the datasets available in its local storage must also cache the $container$ image from the previous execution. In this way, two uncertain factors $T_{vmt}^{d_{t}^{in}}$ and $prov_c$ are eliminated. In this case, the sub-budget for this particular task can be spent well on using the fastest VM type to minimize its execution time. This scenario is always preferred since the retrieval of datasets from $GS$ and downloading the $container$ images through networks has become a well-known overhead that poses a significant uncertainty as its performance also may be degraded over time \cite{6848061}.

If VM has not been found in $VM_{idle}^{input}$, the algorithm finds a set of $VM_{idle}$ that have $container$ deployed. This set of $VM_{idle}^{container}$ may not have the input datasets available as they may have been cleared from VMs local storage. If the set still does not contain the preferred VM, any VM from remaining set of $VM_{idle}$ is chosen. In the last scenario, the overhead of provisioning delay $prov_{vm}$ can be eliminated. It is still better than having to acquire a new VM. Whenever a suitable VM in the resource pool is found, the task then is immediately scheduled on it. If reusing an existing VM is not possible, then the algorithm provisions a new VM of the fastest VM type that can finish the task within its sub-budget. This approach is the last option to schedule a task on the platform.

For better adaptation to uncertainties that come from performance variation and unexpected delays during execution, there is a control mechanism within the algorithm to adjust sub-budget allocations whenever a task finishes execution dynamically. This mechanism defines a spare budget variable that stores the residual sub-budget calculated from the actual cost execution. Whenever a task is finished, the algorithm calculates the actual cost of execution using the Eq. \ref{eq:cost} and redistributes the leftover sub-budget to the unscheduled tasks. If the actual cost exceeds the allocated sub-budget, the shortfall will be taken from the sub-budget of unscheduled tasks. Therefore, the budget update (i.e., budget redistribution) will take place every time a task is finished. In this way, the uncertainties (e.g., performance variation, overhead delays) that befall to a particular task do not propagate throughout the rest of the following tasks. The details of this process are depicted in Algorithm \ref{algo:budgetupdate}.

Regarding the resource provisioning strategy, the algorithm encourages a minimum number of VMs usage by reusing as much as possible existing VMs. The new VMs are only acquired whenever idle VMs are not available due to the high density of the workload to be processed. In this way, the VM provisioning delays overhead can be reduced. As for the deprovisioning strategy, all of the running VMs are monitored every $prov_{int}$ and all VMs that have been idle for more than $threshold_{idle}$ time are terminated as seen in Algorithm \ref{algo:resprov}. The decision to keep or terminate a particular VM should be carefully considered as the cached data within its local storage is one of the valued factors that impact the performance of this algorithm. Therefore, the $prov_{int}$ and $threshold_{idle}$ are configurable parameters that can lead to a trade-off between performance in term of resource utilization and makespan along with the VMs local storage caching policy.

\section{Performance Evaluation}\label{section:performanceeval}

\begin{table}[!t]
	\vspace{-1mm}
	\centering
	\caption{Characteristics of synthetic workflows}
	\label{table:workflows}
	\vspace{-2mm}
	\resizebox{\linewidth}{!}{\begin{tabular}{@{\extracolsep{4pt}} c c c c c@{}}
			\hline \noalign{\vskip 1mm}
			\textbf{Workflow} & \textbf{Parallel Tasks} & \textbf{CPU Hours} &\textbf{I/O Requirements}&\textbf{Peak Memory} \\
			
			\hline \noalign{\vskip 1mm}
			CyberShake&Very High &Very High&Very High&Very High\\
			Epigenome&Medium &Low&Medium&Medium\\
			LIGO&Medium High & Medium&High&High\\
			Montage&High & Low&High&Low\\ 
			SIPHT&Low &Low&Low&Medium\\
			\hline
	\end{tabular}}
\end{table}

To evaluate our proposal, we used five synthetic workflows derived from the profiling of well-known workflows \cite{JUVE2013682} from various scientific areas generated using the WorkflowGenerator tool\footnote{https://confluence.pegasus.isi.edu/display/pegasus/WorkflowGenerator}. The first workflow is CyberShake, that generates synthetic seismograms to differentiate various earthquakes hazards. This earth-science workflow is data-intensive with very high CPU and memory requirements. The second workflow is Epigenome, a Bioinformatics application with CPU-intensive tasks for executing operations that are related to the genome-sequencing research. The third workflow is an astrophysics application called Inspiral--part of the LIGO project--that is used to analyze the data from gravitational waves detection. This workflow consists of CPU-intensive tasks and requires a high memory capacity. The next workflow is Montage. An astronomy application used to produce a sky mosaics image from several different angle sky observation images. Most of the Montage tasks are considered I/O intensive while involving less CPU processing. Finally, we include another Bioinformatics application used to encode sRNA genes called SIPHT. Its tasks have relatively low I/O utilization with medium memory requirements. The resume of these characteristics can be seen in Table \ref{table:workflows}.

The experiments were conducted with various workloads composed of a combination of workflows mentioned above in three different sizes: approximately 50 tasks (small), 100 tasks (medium), and 1000 tasks (large). Each workload contains a different number and various types of workflows that were randomly selected based on a uniform distribution, and the jobs' arrival was modeled as a Poisson process. Every workflow in a workload was assigned a budget that is always assumed sufficient. Budget insufficiency can be managed by rejecting the job from the platform or re-negotiating the budget with the users. This budget was randomly generated based on a uniform distribution from a range of minimum and maximum cost of executing the workflow. The minimum cost was estimated from simulating the execution of all of its tasks in sequential order on the cheapest VM type. On the other hand, the maximum cost was estimated based on the parallel execution of each task on multiple VMs. In this experiment, we used the runtime generated from the WorkflowGenerator for the size measurement of the task.

\begin{table}[!t]
	\vspace{-1mm}
	\centering
	\caption{Configuration of VM types used in evaluation}
	\label{table:vm}
	\vspace{-2mm}
	\resizebox{.975\linewidth}{!}{\begin{tabular}{@{\extracolsep{4pt}} c c c c @{}}
			\hline \noalign{\vskip 1mm}
			\textbf{Name} & \textbf{vCPU (MIPS)} & \textbf{Storage (GB)} &\textbf{Price per second (cent)} \\
			
			\hline \noalign{\vskip 1mm}
			Small&2 & 20&1\\
			Medium&4 & 40&2\\
			Large&8 & 80&4\\ 
			XLarge&16 & 160&8\\
			\hline
	\end{tabular}}
\end{table}

We extended CloudSim \cite{SPE:SPE995} to support the simulation of WaaS cloud platforms. Using CloudSim, we modeled a single IaaS cloud provider that offers a data center within a single availability zone with four VM types that are shown in Table \ref{table:vm}. These four VM types configurations are based on the compute-optimized (c4) instance types offered by the Amazon EC2, where the CPU capacity has a linear relationship with its respective price. We modeled the per-second billing period for leasing the VMs, and for all VM types, we set the provisioning delay to 45 seconds based on the latest study by Ulrich et al. \cite{10.1007/978-3-030-02738-4_19}. On the other side, using the model published by Piraghaj et al. \cite{doi:10.1002/spe.2422}, the container provisioning delay was set to 10 seconds based on the average container size of 600 MB, a bandwidth 500 Mbps, and a 0.4 seconds delay in container initialization.

\begin{figure*}[!t]
	\begin{multicols}{2}
		\centering \includegraphics[trim={3.5cm 0 1cm 0},clip,height=.2\textheight]{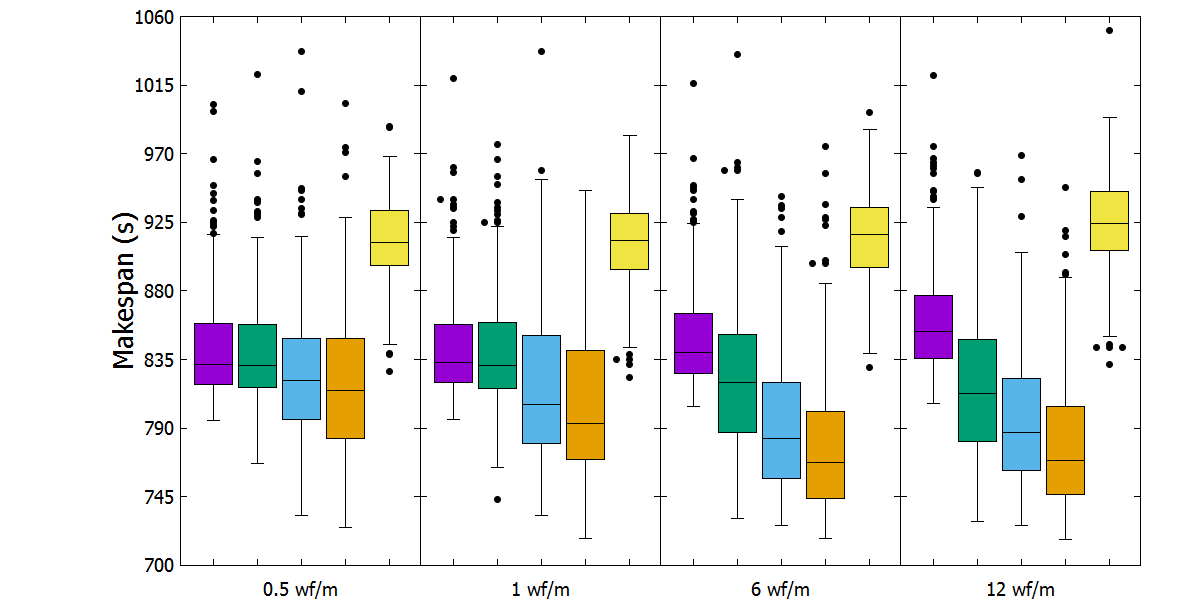}\par
		\subcaption{Makespan of CyberShake}
		\label{fig:cybershake1}
		\centering \includegraphics[trim={3.5cm 0 2cm 0},clip,height=.2\textheight]{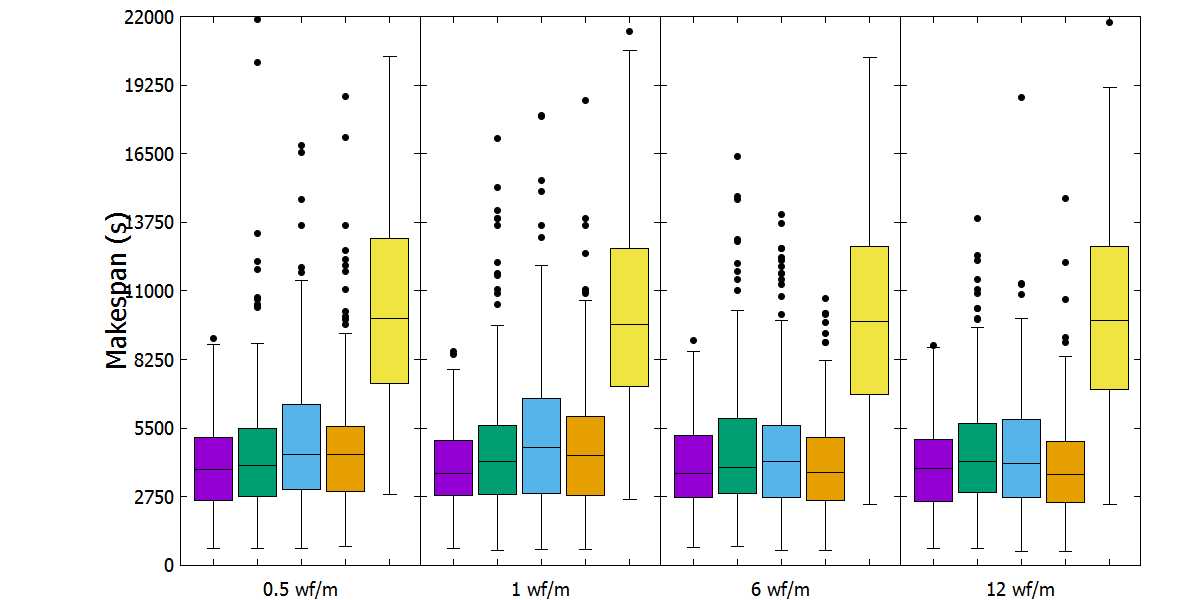}\par
		\subcaption{Makespan of Epigenome}
		\label{fig:epigenomics1}
	\end{multicols}
	\vspace{-7mm}
	\begin{multicols}{2}
		\centering \includegraphics[trim={4cm 0 2cm 0},clip,height=.2\textheight]{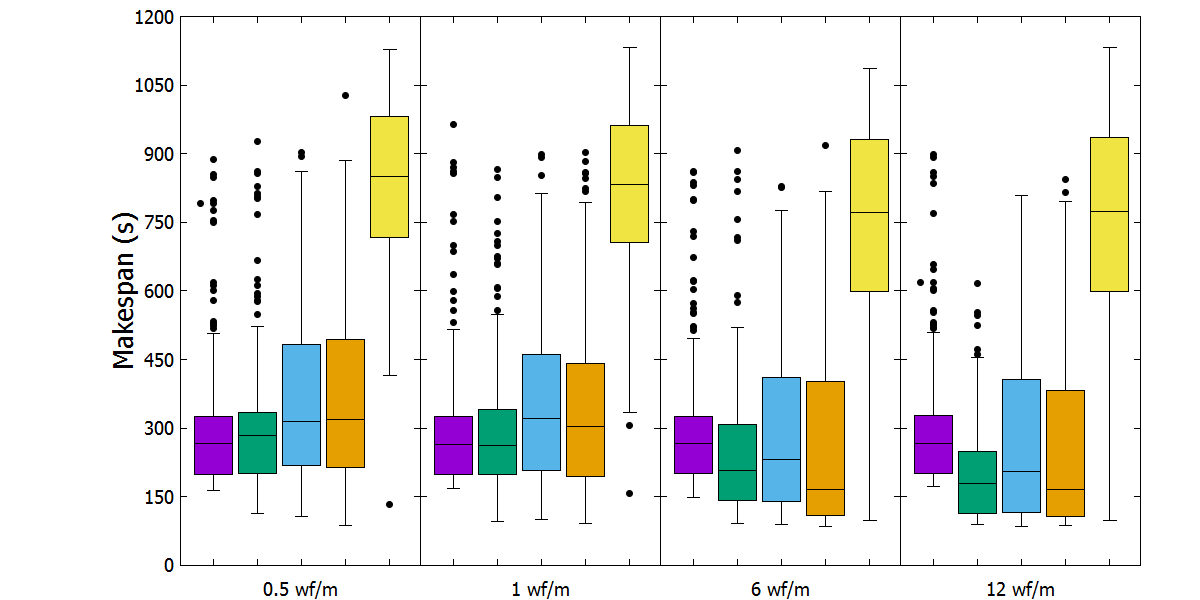}\par
		\subcaption{Makespan of LIGO}
		\label{fig:ligo1}
		\centering \includegraphics[trim={4cm 0 2cm 0},clip,height=.2\textheight]{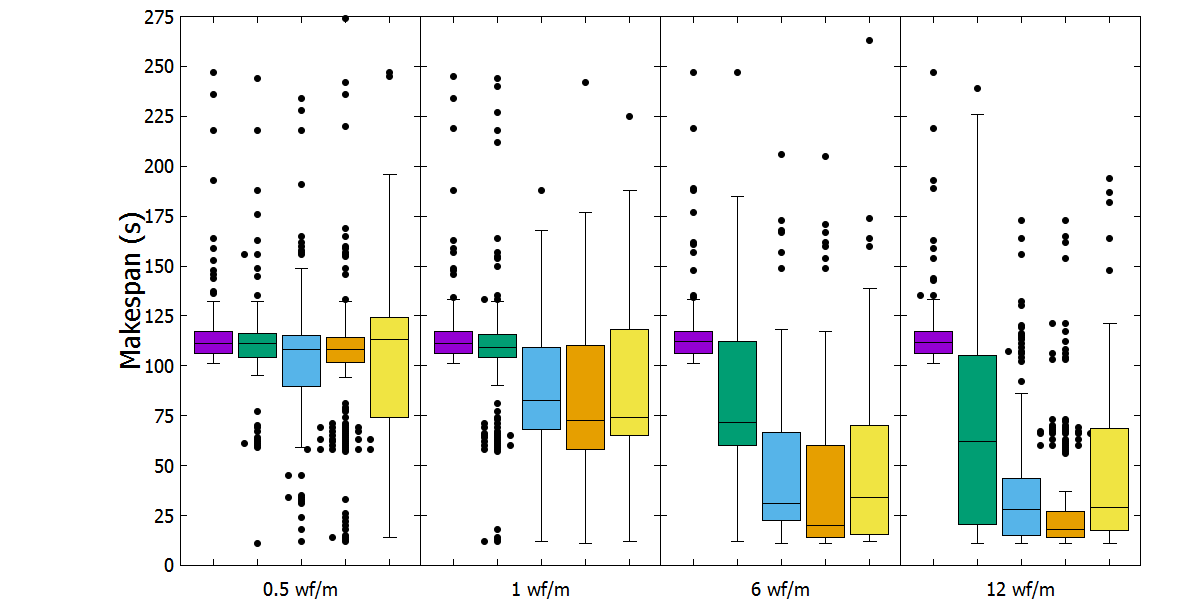}\par
		\subcaption{Makespan of Montage}
		\label{fig:montage1}
	\end{multicols}
	\vspace{-1.5mm}
	\begin{minipage}[c]{1.05\linewidth}
		\centering \includegraphics[trim={4cm 0 2cm 0},clip,height=.2\textheight]{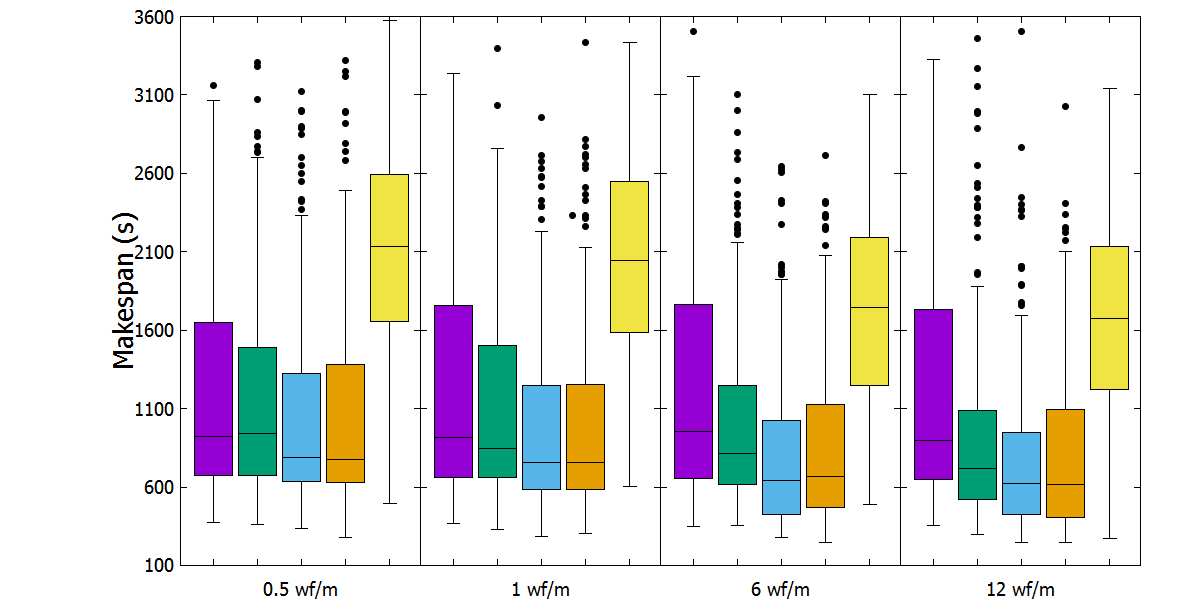}
		\includegraphics[trim={26.5cm 5cm 0 0},clip,height=.19\textheight]{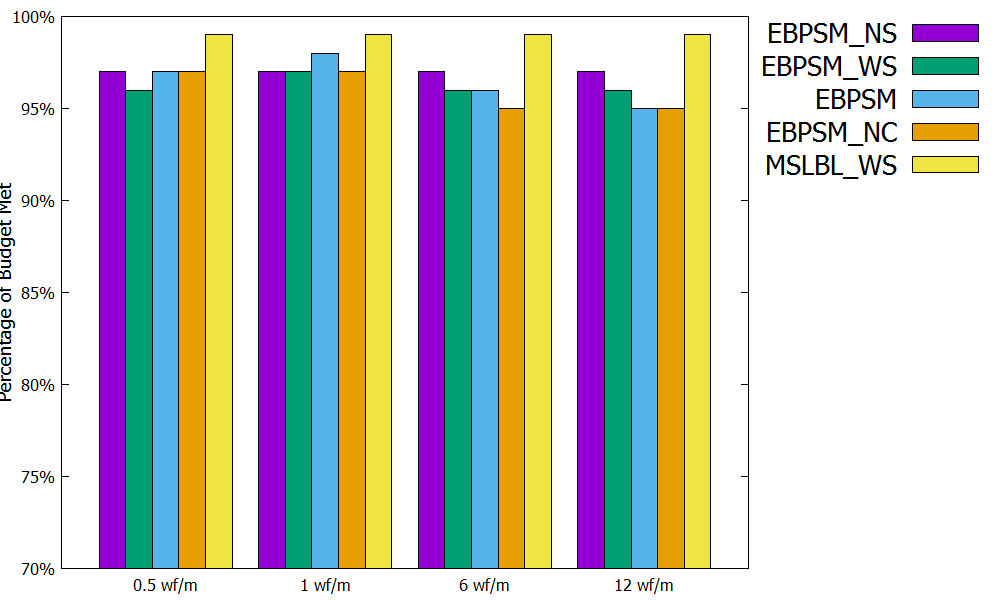}\par
		\subcaption{Makespan of SIPHT}
		\label{fig:sipht1}
	\end{minipage}
	\vspace{-1mm}
	\caption{Makespan of workflows on various workloads with different arrival rate}
	\vspace{-3mm}
\end{figure*}

The CPU and network performance variation were modeled based on the findings by Leitner and Cito \cite{Leitner:2016:PCS:2926746.2885497}. The CPU performance was degraded by a maximum 24\% of its published capacity based on a normal distribution with a 12\% mean and a 10\% standard deviation. Furthermore, the bandwidth available for a data transfer was potentially degraded by at most 19\% based on a normal distribution with a 9.5\% mean and a 5\% standard deviation. In this experiment, as mentioned earlier, each VM was modeled to maintain an $LS_{vmt}$ that stores the cached data produced during the execution based on FIFO policy. The design for a more sophisticated strategy to maintain or terminate the lifetime of cached datasets within an $LS_{vmt}$ is left for future work.

To create a fair baseline for EBPSM, we extended the MSLBL \cite{CHEN20171} algorithm for multiple workflows scheduling (MSLBL\_MW). MSLBL was designed for single workflow execution, so we added a function to handle multiple workflows by creating a pool of arriving workflows where the algorithm then dispatched the ready tasks from all workflows for scheduling. Furthermore, MSLBL assumed that a set of computational resources are available in a fixed quantity all over the scheduling time. Therefore, to cope up with a dynamic environment in WaaS cloud platforms, we added a simple dynamic provisioner for MSLBL\_MW that provisions a new VM whenever there is no existing VMs available. This newly provisioned VM is selected based on the fastest VM type that can be afforded by the sub-budget of a particular task. This dynamic provisioner also automatically terminates any idle VMs to ensure the optimal utilization of the system. Finally, for MSLBL\_MW, we assumed that every VM can contain software configurations for every workflow application type and can be shared between any users in WaaS cloud platforms.

To demonstrate the benefits of our resource-sharing policy, we implemented three additional versions of EBPSM, which are EBPSM\_NS, EBPSM\_WS, and EBPSM\_NC. The EBPSM\_NS does not share the VMs between different users; it is a version of EBPSM that executes each workflow submitted into the WaaS cloud platform in dedicated service, as shown in Figure \ref{fig:indepenttask}. EBPSM\_WS tailors the software configuration of workflow applications in a VM image instead of containers. Therefore, the algorithm allows only tasks from the same workflow application type (e.g., SIPHT with 50 tasks and SIPHT with 100 tasks) that can share the provisioned VMs during the execution. Meanwhile, EBPSM\_NC ignores the use of containers to store the configuration template and naively assumes that each VM can be shared between many users with different requirements. This version was a direct comparable case for MSLBL\_MW. Finally, the \textit{threshold$_{idle}$} for EBPSM\_WS, EBPSM, and EBPSM\_NC was set to 5 seconds. It means the \textit{vm$_{idle}$} is not immediately terminated whenever it goes idle to accommodate the further utilization of the cached data sets within the VM.

\subsection{To Share or Not To Share}

The purpose of this experiment is to evaluate the effectiveness of our proposed resource-sharing policy regarding its capability to minimize the workflows' makespan while meeting the soft limit budget. In this scenario, we evaluated EBPSM and its variants against MSBLBL\_MW under four workloads with the different arrival rate of 0.5, 1, 6, and 12 workflows per minute. Each workload consists of 1000 workflows with approximately 170 thousand tasks of various size (e.g., small, medium, large) and different workflow's type generated randomly based on a uniform distribution. The arrival rate for these four workloads represents the density of workflows' arrival in the WaaS cloud platform. The arrival of 0.5 workflows per minute represents a less occupied platform, while the arrival of 12 workflows per minute models a busier system in handling the workflows.

\begin{figure*}[!t]
	\begin{multicols}{3}
		\begin{center}
			\includegraphics[height=.145\textheight]{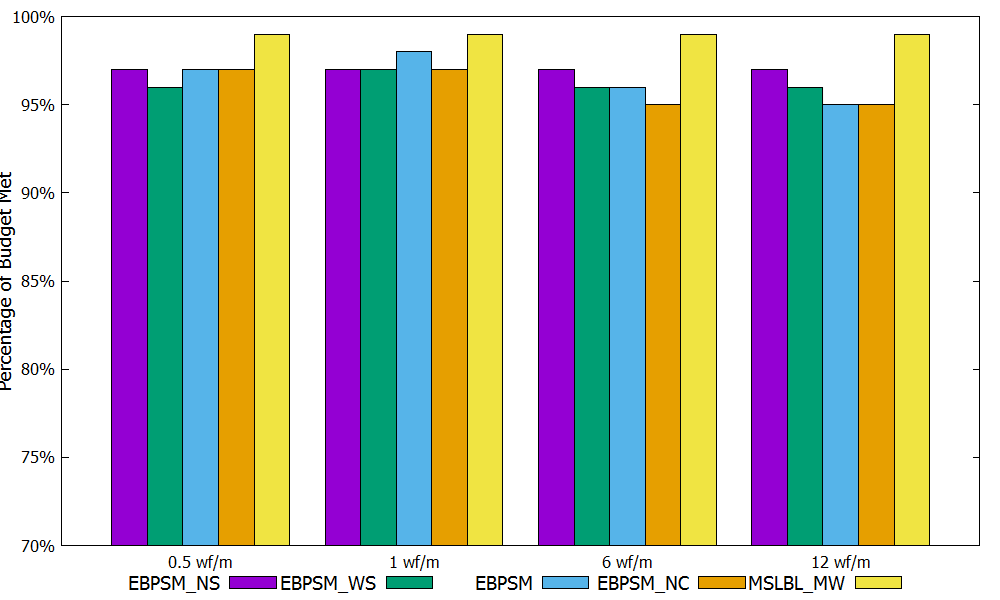}\par
			\subcaption{Percentage of budget met}
			\label{fig:budget1}
		\end{center}
		\begin{center}
			\includegraphics[height=.145\textheight]{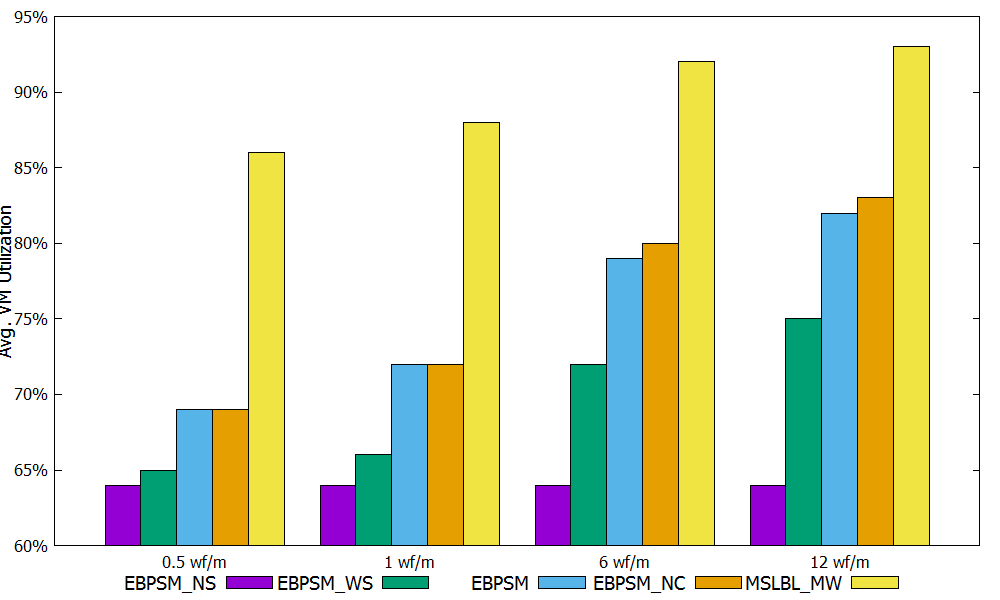}\par
			\subcaption{Avg. VM utilization}
			\label{fig:vmutil}
		\end{center}
		\begin{center}
			\centering \includegraphics[height=.1475\textheight]{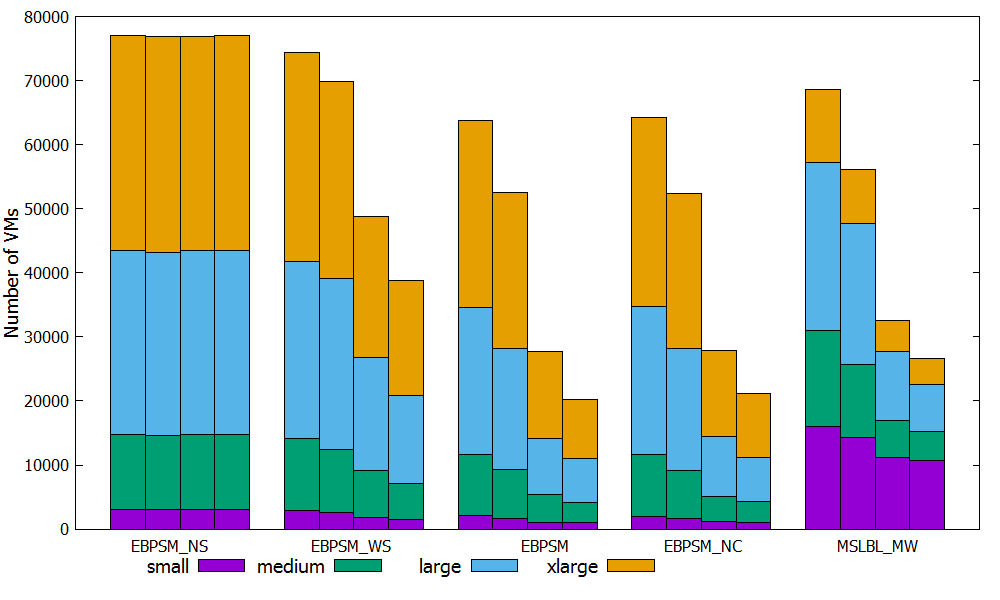}\par
			\subcaption{Number of VMs}
			\label{fig:vmused}
		\end{center}
	\end{multicols}
	\vspace{-4mm}
	\caption{Percentage of budget met and VM usage on various workloads with different arrival rate}
	\vspace{-4mm}
\end{figure*}

Figures \ref{fig:cybershake1}, \ref{fig:epigenomics1}, \ref{fig:ligo1}, \ref{fig:montage1}, and \ref{fig:sipht1} depict the makespan achieved for CyberShake, Epigenome, LIGO, Montage, and SIPHT workflows respectively. EBPSM\_NS that represents the traditional non-shared cloud resources paradigm, shows almost no difference in the algorithm's performance across different arrival rate. This version of the algorithm serves each of the workflows in dedicated and isolated resources. Therefore, it can maintain a similar performance for all four workloads. However, on the other hand, EBPSM\_NS shows the lowest percentage of average VM utilization due to this non-sharing policy, as seen in Figure \ref{fig:vmutil}.

In contrast to EBPSM\_NS, the other three versions (e.g., EBPSM\_WS, EBPSM, EBPSM\_NC) exhibit a performance improvement along with the increasing density of the workloads. This further makespan minimization is the result of (i) the elimination of data transfer overhead between tasks' execution and (ii) the utilization of inevitable scheduling gaps between tasks' execution. In the case of data transfer elimination, we can observe that the improvement is relatively not significant for Epigenome workflows where the CPU processing takes the biggest portion of the execution time instead of I/O and data movement. On the other hand, the most significant improvement can be observed from the data-intensive workflows such as Montage and CyberShake applications. Furthermore, the superiority of EBPSM and EBPSM\_NC over EBPSM\_WS both in makespan and average VM utilization shows a valid argument for the schedule gaps utilization case. From the following result, we can conclude that the utilization of idle gaps between users from different workflow types can further minimize the makespan.

\begin{table}[!b]
	\vspace{-1mm}
	\centering
	\caption{Cost/budget ratio for EBPSM budget violated cases}
	\label{table:budgetratio}
	\vspace{-2mm}
	\label{table:percentilearr}
	\resizebox{.95\linewidth}{!}{\begin{tabular}{@{\extracolsep{4pt}} c c c c c@{}}
			\hline \noalign{\vskip 1mm}
			\textbf{Percentile} & \textbf{0.5 wf/m} & \textbf{1 wf/m} &\textbf{6 wf/m} & \textbf{12 wf/m} \\
			
			\hline \noalign{\vskip 1mm}
			10th&1.005&1.004&1.017&1.005\\
			30th&1.017&1.018&1.032&1.023\\
			50th&1.026&1.030&1.051&1.052\\ 
			70th&1.046&1.053&1.065&1.069\\
			90th&1.072&1.083&1.121&1.107\\
			\hline
	\end{tabular}}
	\vspace{-2mm}
\end{table}

The naive assumption that every VM can be shared between any users in the platform explains the lower makespan and the higher utilization produced by EBPSM\_NC compared to EBPSM. Container usage generates additional initialization delays that affect EBPSM performance. However, the difference between them is marginal, and EBPSM still exhibits its superiority to the other versions.

\begin{figure*}[!t]
	\vspace{-2.5mm}
	\begin{multicols}{3}
		\centering \includegraphics[height=.155\textheight]{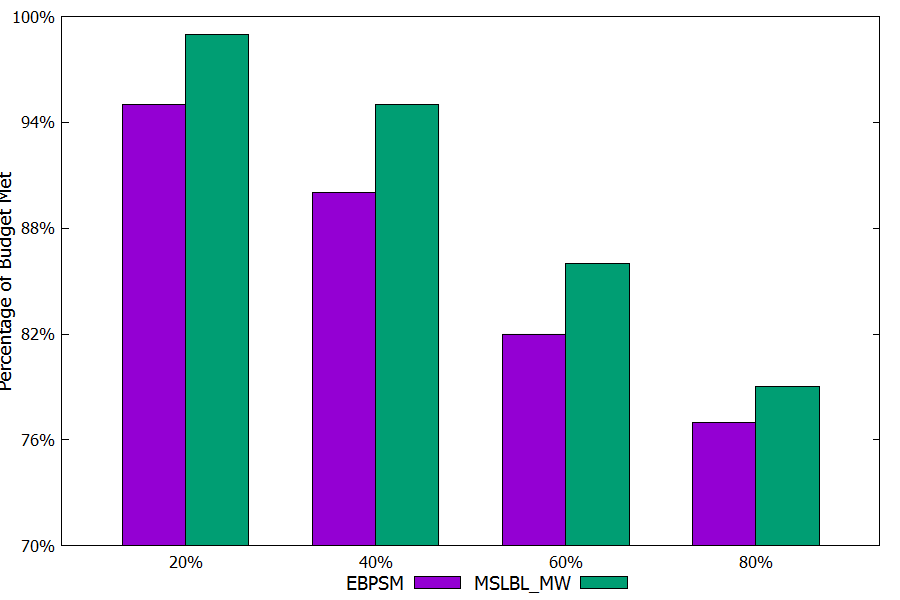}\par
		\subcaption{Percentage of budget met}
		\label{fig:budget2}
		\centering \includegraphics[trim={2.5cm 0 1.5cm 0},clip,height=.1575\textheight]{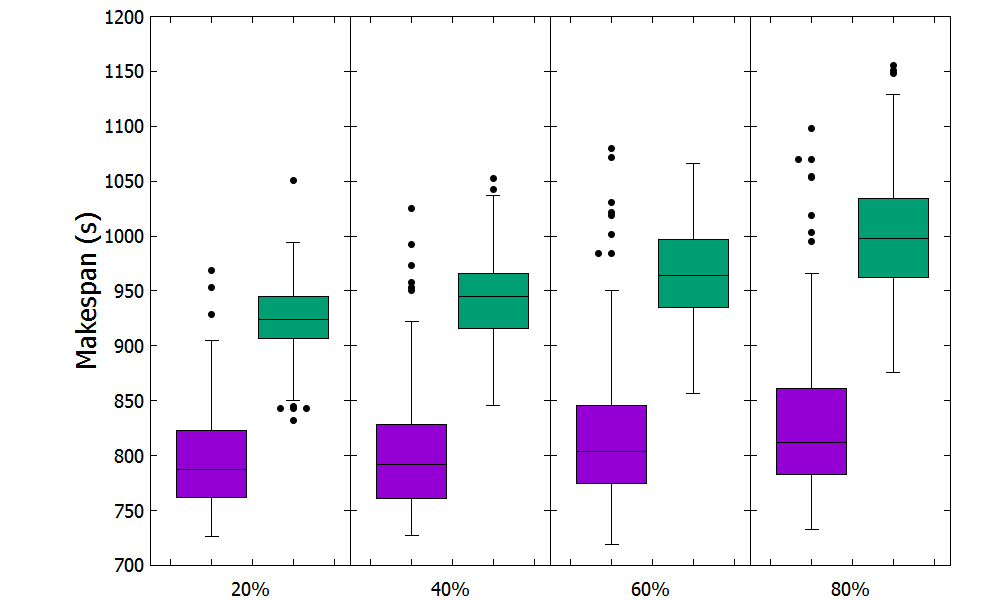}\par
		\subcaption{Makespan of CyberShake}
		\label{fig:cybershake2}
		\centering \includegraphics[trim={2.5cm 0 1.5cm 0},clip,height=.1575\textheight]{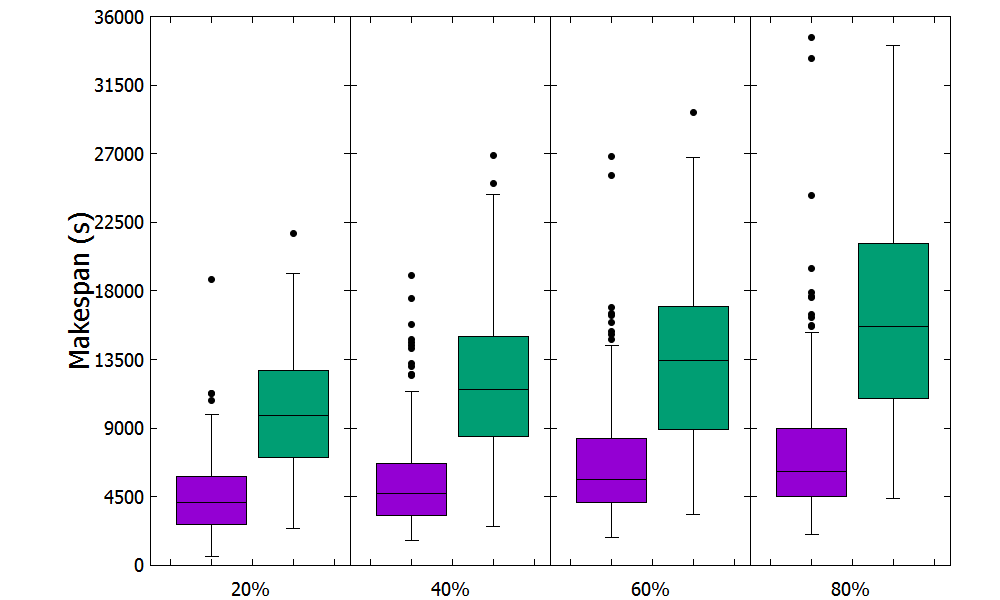}\par
		\subcaption{Makespan of Epigenome}
		\label{fig:epigenomics2}
	\end{multicols}
	\vspace{-6mm}
	\begin{multicols}{3}
		\centering \includegraphics[trim={2.5cm 0 1.5cm 0},clip,height=.1575\textheight]{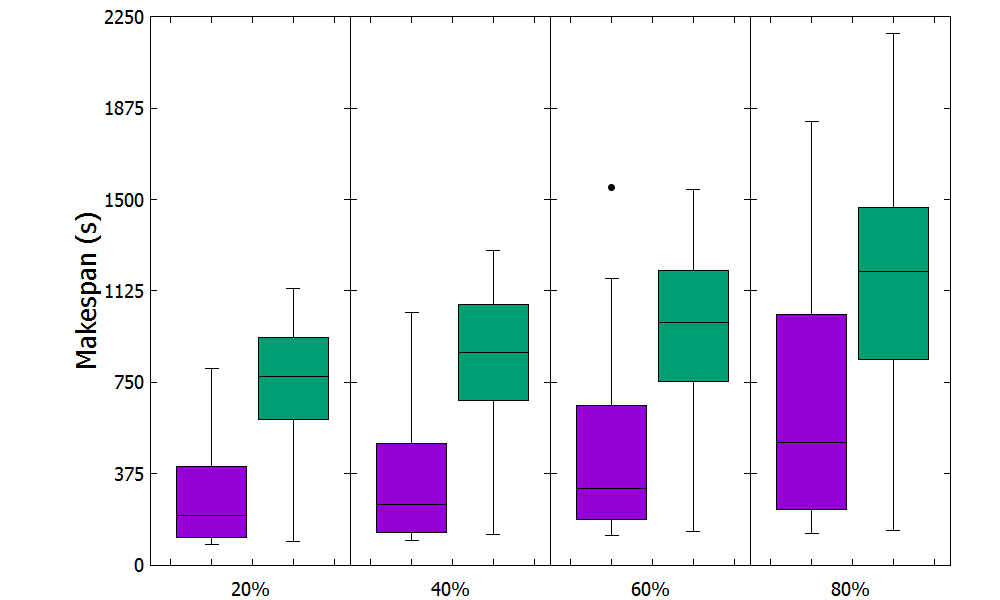}\par
		\subcaption{Makespan of LIGO}
		\label{fig:ligo2}
		\centering \includegraphics[trim={2.5cm 0 1.5cm 0},clip,height=.1575\textheight]{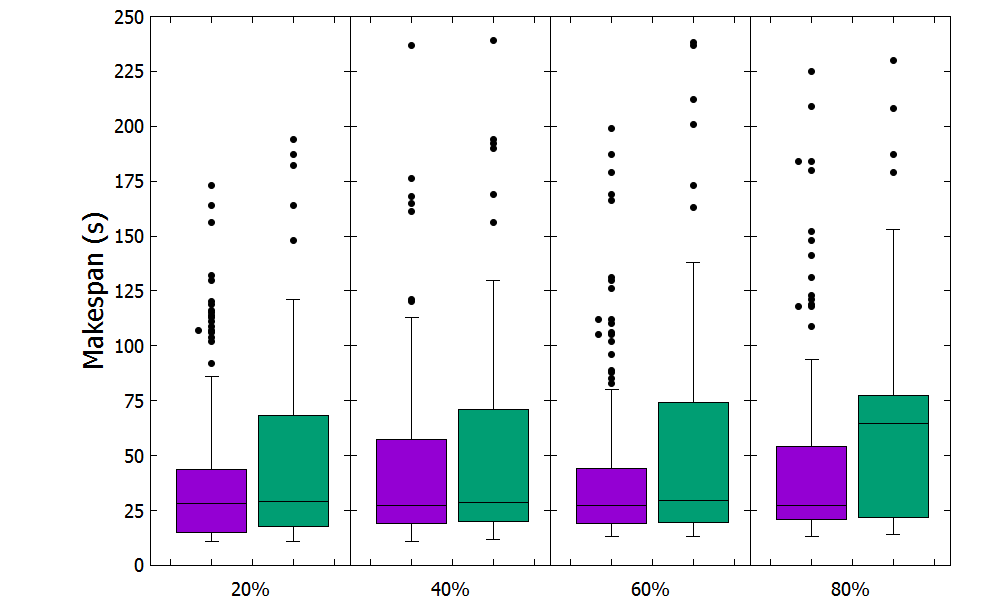}\par
		\subcaption{Makespan of Montage}
		\label{fig:montage2}
		\centering \includegraphics[trim={2.5cm 0 1.5cm 0},clip,height=.1575\textheight]{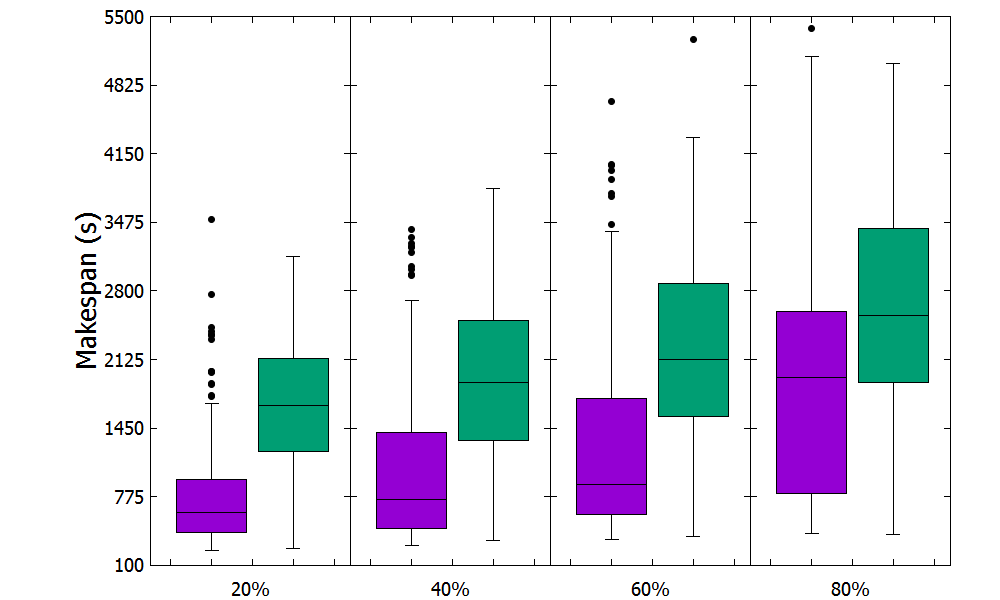}\par
		\subcaption{Makespan of SIPHT}
		\label{fig:sipht2}
	\end{multicols}
	\vspace{-4mm}
	\caption{Percentage of budget met and makespan of workflows on various CPU performance degradation}
	\label{fig:makespanperfdegrad}
	\vspace{-1mm}
\end{figure*}

We observe that in four out of five workflows cases, all versions of EBPSM overthrow MSLBL\_MW regarding the makespan achievement. This result comes from the different strategies of both algorithms in distributing the budget and avoiding the constraint's violation, which implies the type of VMs they provisioned. EBPSM prioritizes the budget allocation to the earlier tasks and leases the fastest VM type as much as possible. This approach is based on the idea that the following children tasks can utilize already provisioned VMs while maintaining the capability of meeting the budget by updating the allocation based on the actual tasks' execution.

On the other hand, MSLBL\_MW allocates the budget based on the budget level factor, which creates a safety net by provisioning the VM that costs somewhere between the minimum and maximum execution cost of a particular task. In this way, MSLBL\_MW reduces the possibility of budget violation at the budget distribution phase. These two different approaches result in the different number of VM types used during the execution as can be seen from Figure \ref{fig:vmused}. MSLBL\_MW leases a lower number of faster VM types compared to EBPSM for all cases. The only case where the performance of MSLBL\_MW is relatively equal to EBPSM is in Montage workflow where the tasks are relatively short in CPU processing time, while the significant portion of their execution time takes place in the data movement. In this Montage case, the decision to lease which kind of VM type does not significantly affect the total makespan.

In Figure \ref{fig:budget1}, we can see that all of the algorithms can achieve at least 95\% of the budget meeting for all cases. The margin between MSLBL\_MW and the four versions of EBPSM was never wider than 4\%. MSLBL\_MW is superior to EBPSM regarding the average VM utilization. This result is caused by the difference in the VM deprovisioning policy. MSLBL\_MW eliminates any VMs as soon as they become idle, while EBPSM delays the elimination in the hope of further utilization and cached data for the following tasks on that particular idle VM. In this case, the configurable settings of \textit{threshold$_{idle}$} value may affect the VM utilization. However, from these two different approaches, a significant margin of makespan between MSLBL\_MW and EBPSM can be observed in most cases.

\begin{figure}[!b]
	\centering
	\begin{subfigure}[b]{0.474\linewidth}
		\includegraphics[width=\linewidth]{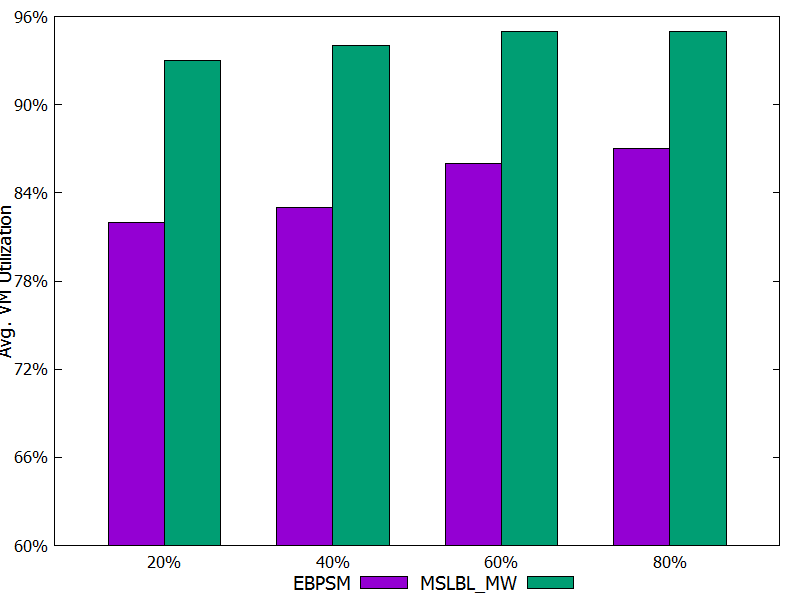}
		\subcaption{Avg. VM Utilization}    
		\label{fig:vmutil2}
	\end{subfigure}    
	\quad
	\begin{subfigure}[b]{0.474\linewidth}
		\includegraphics[width=\linewidth]{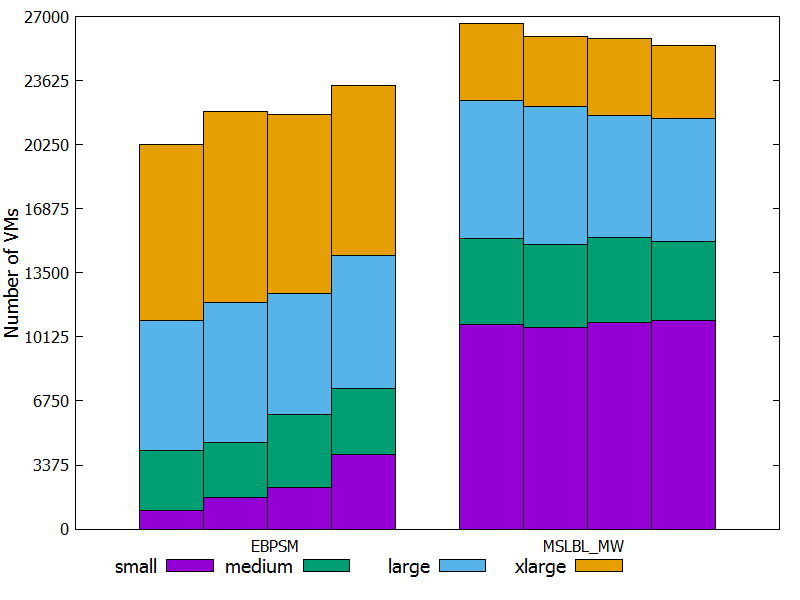}
		\subcaption{Number of VMs}
		\label{fig:vmused2}
	\end{subfigure}
	\caption{VM usage on various CPU performance degradation}
	\vspace{-3mm}
\end{figure}

We captured the cases where EBPSM failed to meet the budget and show the result in Table \ref{table:budgetratio} to better understand the EBPSM performance. From the table, we can see that the cost/budget ratio for 90\% of the EBPSM budget violation cases are lower than 1.12. This ratio means that the additional cost produced from these violations is never higher than 12\%. This percentage is relatively small and may be caused by an extreme case of CPU and network performance degradation. It is not possible to eliminate the negative impact of these uncertainties in such a dynamic environment.

\subsection{Performance Degradation Sensitivity}

Adapting to performance variability is an essential feature for schedulers in multi-tenant dynamic environments. This ability ensures the platform can quickly recover from unexpected events that may occur at any given time, hence preventing a snowball effect that negatively impacts subsequent executions. In this experiment, we evaluate the sensitivity of EBPSM and MSLBL\_MW--on the default environment--to CPU performance degradation by analyzing the percentage of budgets met, makespan, average VM utilization, and the number of VMs used on four different scenarios. We model CPU performance degradation using a normal distribution with 1\% variance and different average and maximum values. The average value is defined as half of the maximum of the CPU performance degradation which ranges from 20\% to 80\%.

\begin{figure*}[!t]
	\vspace{-2.5mm}
	\begin{multicols}{3}
		\centering \includegraphics[trim={0cm 0 0cm 0},clip,height=.1575\textheight]{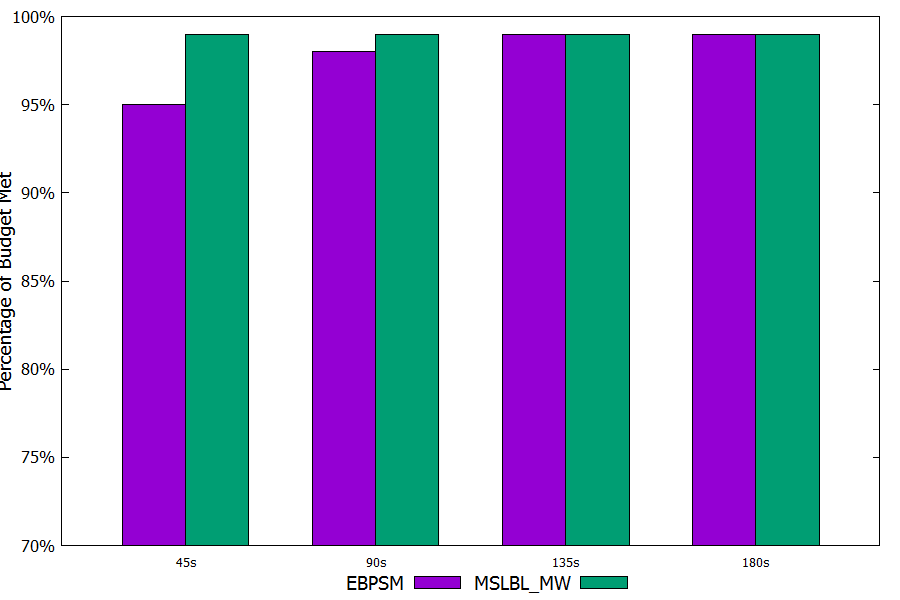}\par
		\subcaption{Percentage of budget met}
		\label{fig:budget3}
		\centering \includegraphics[trim={2.5cm 0 1.5cm 0},clip,height=.1575\textheight]{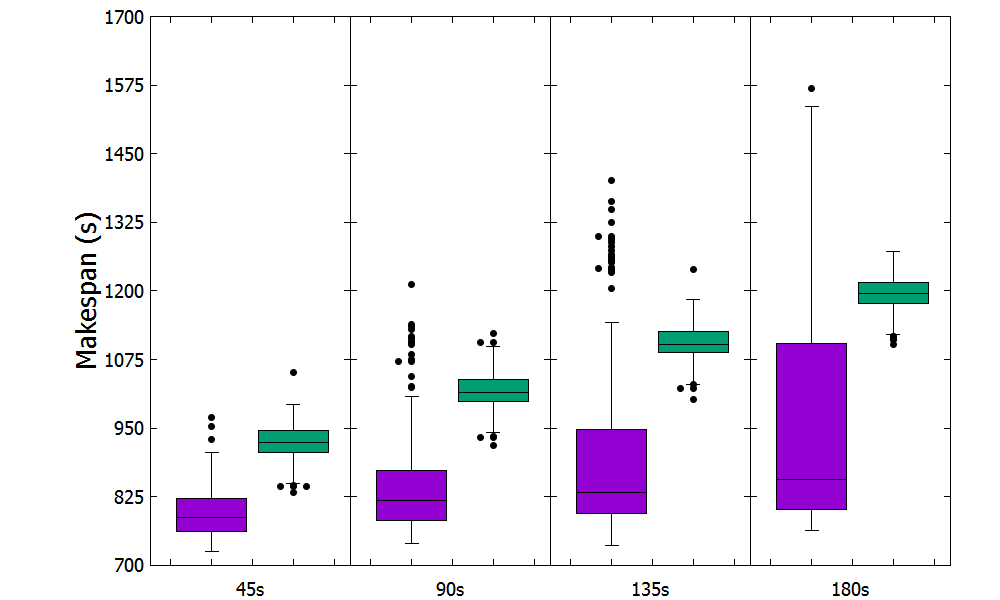}\par
		\subcaption{Makespan of CyberShake}
		\label{fig:cybershake3}
		\centering \includegraphics[trim={2.5cm 0 1.5cm 0},clip,height=.1575\textheight]{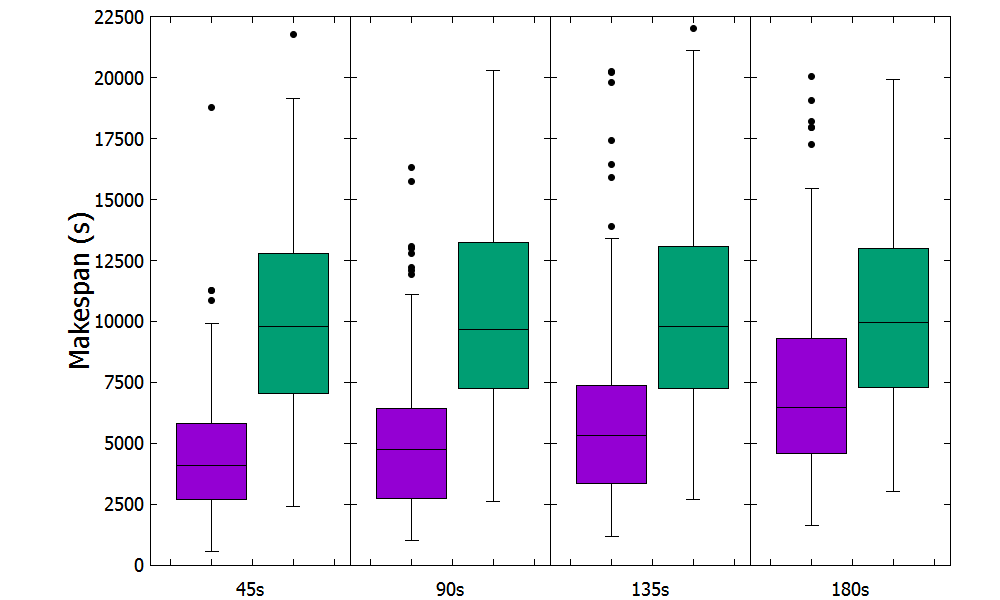}\par
		\subcaption{Makespan of Epigenome}
		\label{fig:epigenomics3}
	\end{multicols}
	\vspace{-6mm}
	\begin{multicols}{3}
		\centering \includegraphics[trim={2.5cm 0 1.5cm 0},clip,height=.1575\textheight]{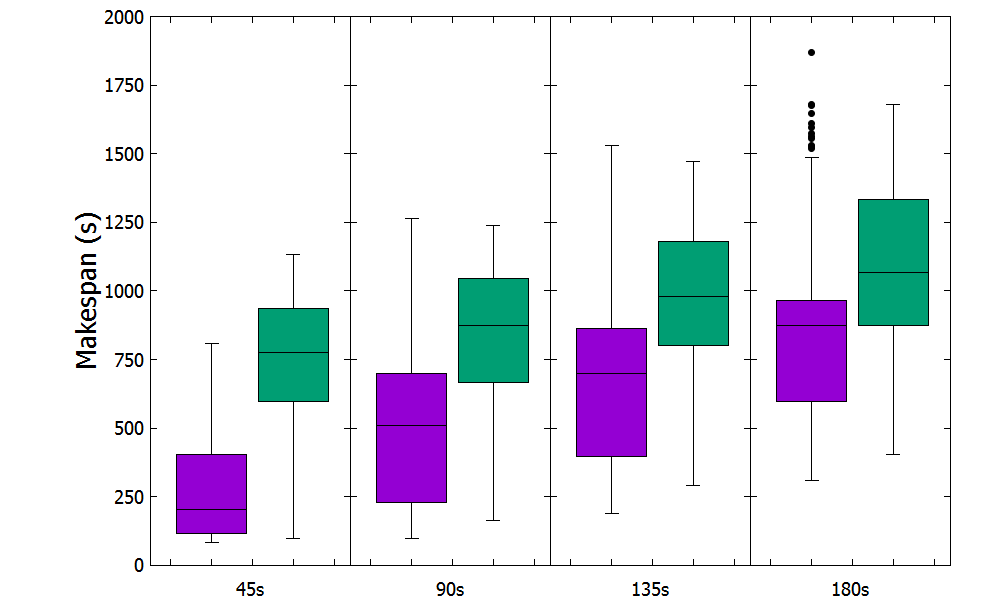}\par
		\subcaption{Makespan of LIGO}
		\label{fig:ligo3}
		\centering \includegraphics[trim={2.5cm 0 1.5cm 0},clip,height=.1575\textheight]{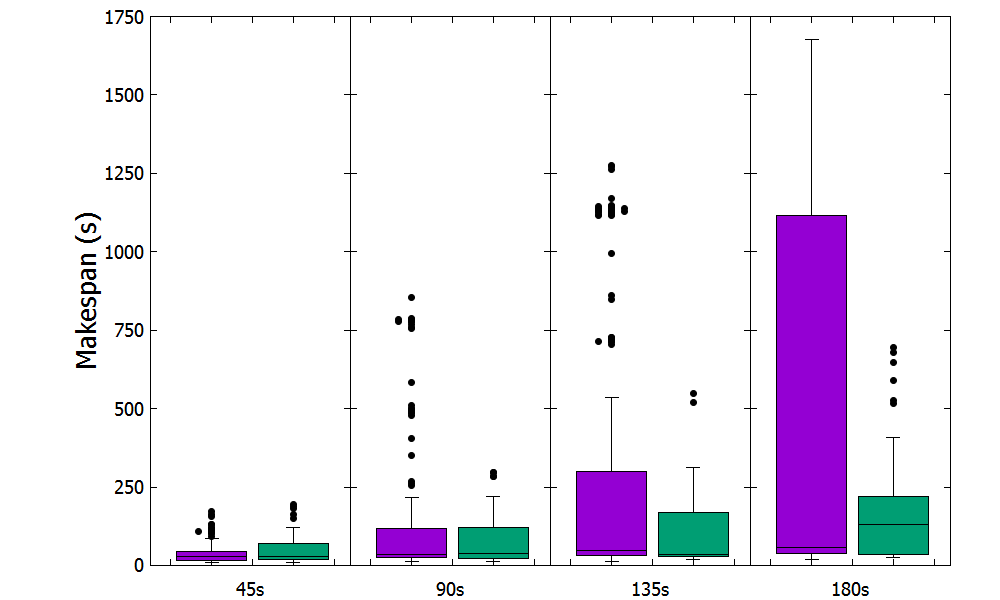}\par
		\subcaption{Makespan of Montage}
		\label{fig:montage3}
		\centering \includegraphics[trim={2.5cm 0 1.5cm 0},clip,height=.1575\textheight]{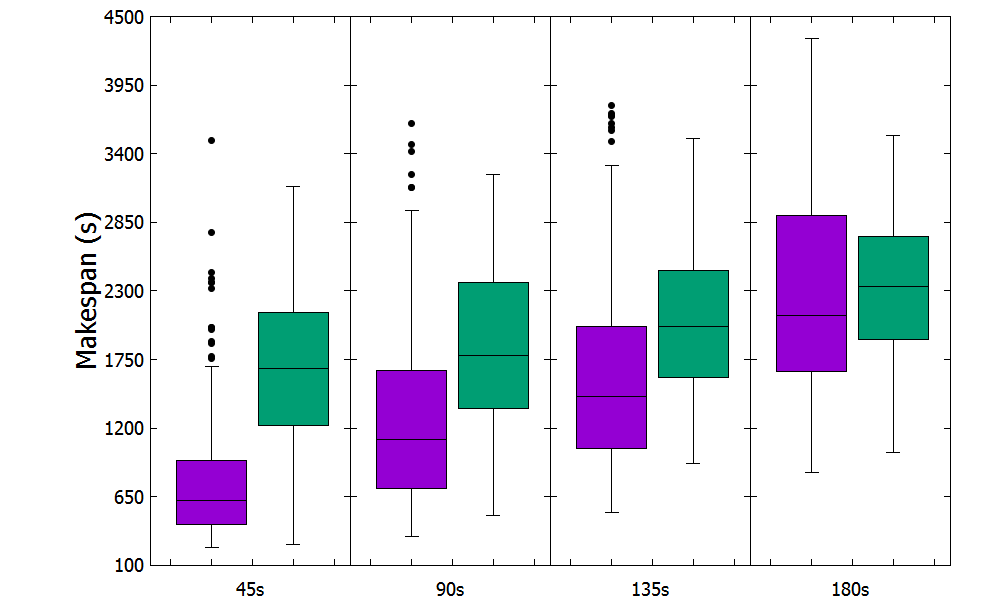}\par
		\subcaption{Makespan of SIPHT}
		\label{fig:sipht3}
	\end{multicols}
	\vspace{-3mm}
	\caption{Percentage of budget met and makespan of workflows on various VM provisioning delay}
	\label{fig:makespanvmprov}
	\vspace{-3mm}
\end{figure*}

All algorithms are significantly affected by CPU performance degradation as their percentage of budget met decreases along with the increased maximum degradation value. However, MSLBL\_MW suffers the most as its performance margin with EBPSM is getting smaller, as seen in Figure \ref{fig:budget2}. This suffering also can be observed from Figures \ref{fig:cybershake2}, \ref{fig:epigenomics2}, \ref{fig:ligo2}, \ref{fig:montage2}, \ref{fig:sipht2}. The increasing makespan as a response to the performance degradation for EBPSM was relatively lower than its effect on MSLBL\_MW. EBPSM can perform better than MSLBL\_MW because of its capability to adapt the changes by evaluating a particular task's execution right after it was finished.

Meanwhile, MSLBL\_MW only relies on the spare budget from its safety net of budget allocation that limits the number of faster VM types at the budget distribution phase. When the maximum degradation value increases, there is no extra budget left from this safety net. Hence, MSLBL\_MW ability to meet the budget drops faster than EBPSM. This reason is also in line with the average VM utilization results where MSLBL\_MW cannot increase VMs utilization as it reaches the top limit of its capabilities, as seen in Figure \ref{fig:vmutil2}.

\begin{figure}[!b]
	\begin{subfigure}[b]{0.474\linewidth}
		\includegraphics[width=\linewidth]{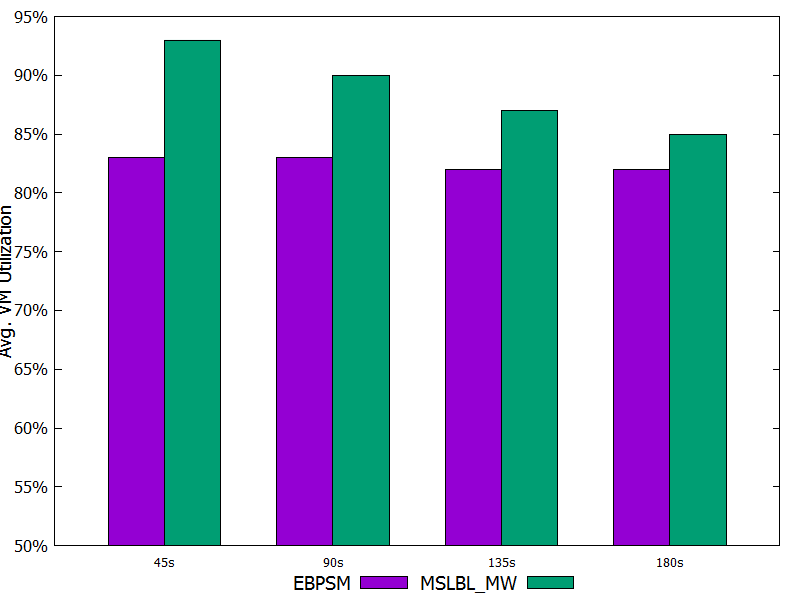}
		\subcaption{Avg. VM Utilization}    
		\label{fig:vmutil3}
	\end{subfigure}    
	\quad
	\begin{subfigure}[b]{0.474\linewidth}
		\includegraphics[width=\linewidth]{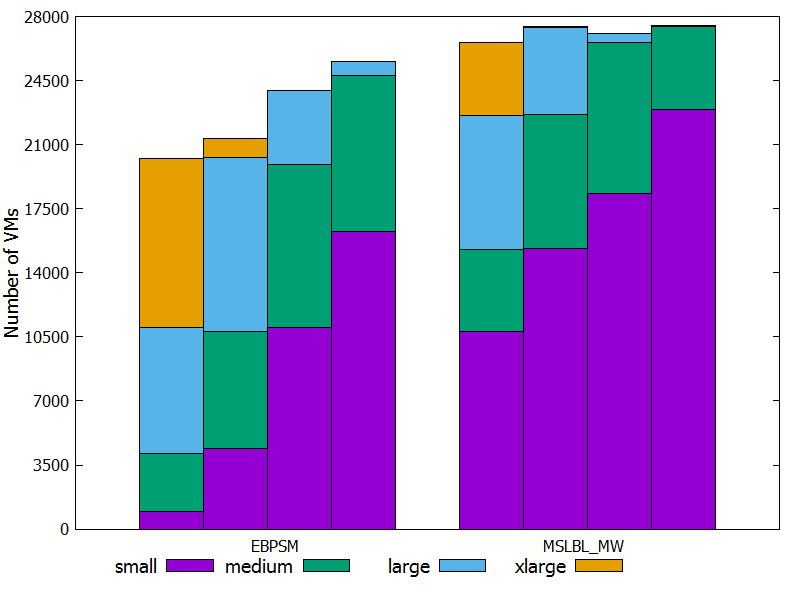}
		\subcaption{Number of VMs}
		\label{fig:vmused3}
	\end{subfigure}
	\caption{VM usage on various CPU provisioning delay}
	\vspace{-3mm}
\end{figure}

Another perspective can be observed from the number of VMs used by each algorithm, as depicted in Figure \ref{fig:vmused2}. EBPSM recovers from the CPU performance degradation by continually increasing the number of slower VM types, while maintaining as much as possible the faster VM types. This adaption is being made through the budget update process after a task finished being executed. Therefore, the earlier tasks are still able to be scheduled onto faster VMs, while the children tasks in the tail of the workflow are inevitably allocated to the slower ones due to the budget left from the recovering process. Different behavior is observed from MSLBL\_MW usage of VMs. We cannot see any significant effort to recover from the number of VM type used. We argue that MSLBL\_MW's way to deal with CPU performance degradation is by highly relying on the safety net of budget allocation from the budget distribution phase.

\subsection{VM Provisioning Delay Sensitivity}

A large volume of arriving workflows results in a vast number of tasks in the scheduling queue to be processed at a given time. Regardless of the effort put into minimizing the number of VMs used by sharing and re-using already provisioned VMs, provisioning a large number of VMs is inevitable in WaaS cloud platform. This provisioning will become a problem if the scheduler is not designed to handle the uncertainties from delays in acquiring the VMs. In this experiment, we study the sensitivity of EBPSM and MSLBL\_MW algorithms--on the default environment--toward VM provisioning delay by analyzing the percentage of budget met, makespan, average VM utilization, and the number of VMs used on four different VM provisioning delay scenarios. The delays range from 45 to 180 seconds.

An interesting point can be observed from Figure \ref{fig:budget3} that shows the percentage of budget met with the algorithms. All of the algorithms perform well, facing increasing VM provisioning delays. However, it comes with a greater trade-off of the workflows' makespan as seen in Figures \ref{fig:cybershake3}, \ref{fig:epigenomics3}, \ref{fig:ligo3} \ref{fig:montage3}, and \ref{fig:sipht3}, especially for the workflow applications that highly depend on CPU processing. We can observe that the spectrum for Montage and CyberShake makespan in EBPSM that results for 180 seconds delays is quite wide although its overall performance is still superior to MSLBL\_MW. In contrast to this situation, the makespan spectrum for MSLBL\_MW for those two workflows is very narrow.

\begin{figure*}[!t]
	\begin{minipage}[c]{\linewidth}
		\centering \includegraphics[trim={0 0 0.5cm 0},clip,height=.215\textheight]{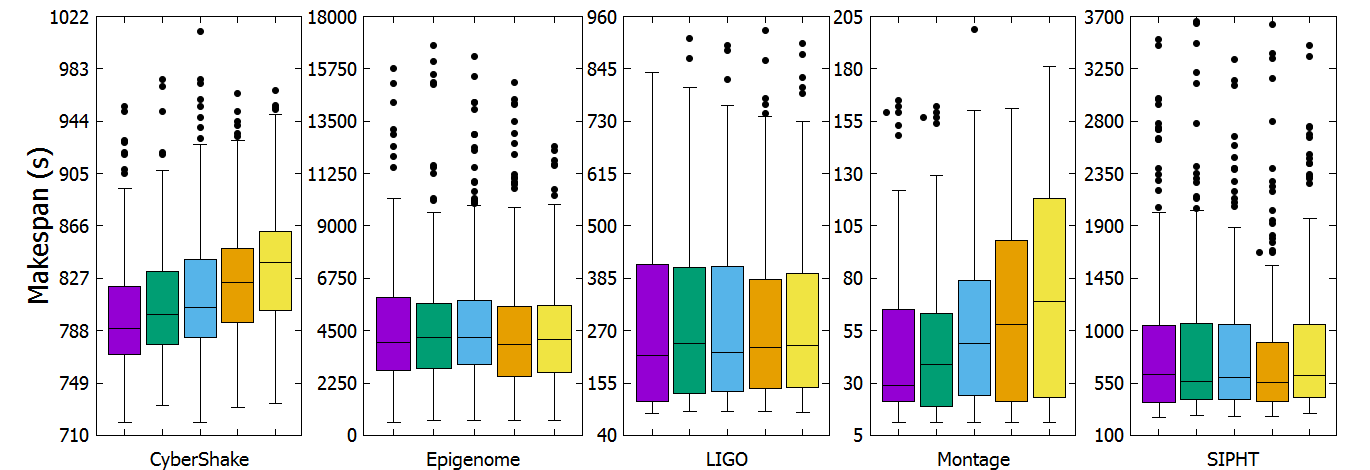}
		\includegraphics[trim={30.15cm 5cm 0 0},clip,height=.205\textheight]{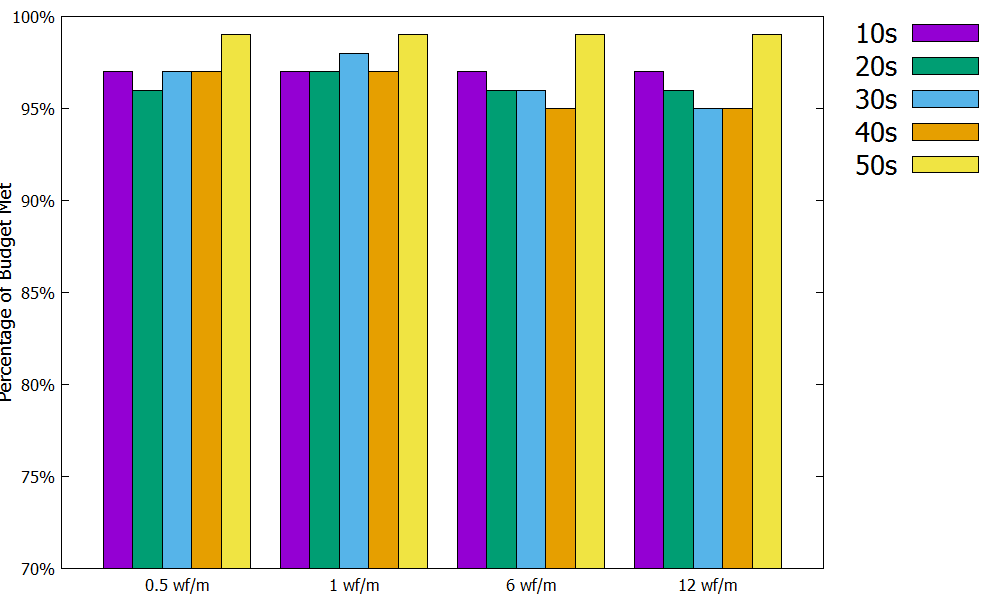}\par
	\end{minipage}
	\vspace{-1mm}
	\caption{EBPSM performance on various container initiating delay}
	\label{fig:contdelay}
	\vspace{-3mm}
\end{figure*}

One of the reasons that may cause this variation in the spectrum of EBPSM's makespan is the variation of the number of shared VMs and the VM type used during the execution. Notably, in the higher VM provisioning delays scenarios. On the other hand, MSLBL\_MW adopts a simple sharing policy. This algorithm does not care whether a particular VM contains a cached data for a future scheduled task or not; it just merely scheduled the task to any idle VM available. In this way, the algorithm can reduce the variation of the VM type used during the execution.

When EBPSM leases faster VMs for the earlier tasks, the VM provisioning delays increase the actual cost for those VMs. Therefore, the algorithm cannot afford those already provisioned VMs if the budget left for the following tasks is not enough to provision the same VM type. In this case, EBPSM must provision a new VM with a slower type. Hence, a workflow with a tight budget will suffer unexpected longer makespan. This condition happens because the algorithm tries to allocate a faster VM type for the earlier tasks but turns out that the budget left is not sufficient. Therefore, to avoid the budget violation, the algorithm automatically recovers during the budget update phase by provisioning a new slower VM type for that particular task. Figure \ref{fig:vmused3} confirms this situation where EBPSM uses a broader variety of VM types compared to MSLBL\_MW. However, in general, the VM provisioning delays does not directly affect the average VM utilization as the VM initiating process is not counted through the total utilization, but still, these delays are being charged. So that is why the delay profoundly affects the total budget. Finally, this analysis does not hinder the fact that EBPSM, in general, is still superior to MSLBL\_MW in terms of makespan for most of the cases.

\subsection{Container Initiating Delay Sensitivity}

Various applications consist of a different number of tasks and software dependencies that may impact the size of the container images. Therefore, in this experiment, we study the sensitivity of EBPSM algorithm--on the default environment--toward container initiating delay by analyzing the percentage of budget met and makespan with five different container initiating delays. The delays range from 10 to 50 seconds.

In this experiment, we found that the periodic budget correction of EBPSM can maintain budget-met compliance of 94-95\% in all scenarios. Therefore, we do not plot the budget-met percentage and present the impact to the makespan instead. From Figure \ref{fig:contdelay}, we can see that the container initiating delay affects the workflows with a high number of tasks and I/O requirements. In this case, CyberShake and Montage makespan increases along with the delays, while the results for the other workflows are relatively indistinguishable.

From these two experiments, we can observe that the container initiating delays may have a higher performance impact to the workflows with a high number of parallel tasks. This condition may be caused by the necessity of deploying a high number of container images to cater to the parallel execution. In this case, a decision to store particular container images of these workflows must be carefully considered. Therefore, a more sophisticated strategy of maintaining container images locality is a crucial aspect to be explored.

\subsection{Experimental Validation}

We have implemented the EBPSM scheduling algorithm in the WaaS cloud platform prototype which extended the functionality of Cloudbus Workflow Engine \cite{RODRIGUEZ2017367} to support multiple workflows execution. We deployed the platform and utilized the Amazon EC2 cloud resources to carry out similar multiple workflows execution scenarios of bioinformatics workflow applications.

Given the experimental settings, the real system evaluation presented an actual impact of the overhead and performance variation of cloud resources—the results shown the fluctuated makespan and budget spending which confirm the dynamicity of cloud computing environments. Overall, the performance evaluation of this platform exhibited similar results to the one presented in this work as a simulated environment. The implementation of the EBPSM algorithm in the WaaS cloud platform prototype along with the performance evaluation can be referred to the report by Hilman et al. \cite{HILMAN2020WAAS}

\section{Conclusions and Future Work}\label{section:conclusion}

The growing popularity of scientific workflows deployment in clouds drives the research on multi-tenant platforms that provides utility-like services for executing workflows. As well as any other multi-tenant platform, this workflow-as-a-service (WaaS) cloud platform faces several challenges that may impede the system effectiveness and efficiency. These challenges include the handling of a continuously arriving workload of workflows, the potential of system inefficiency from inevitable idle time slots from workflows' tasks dependency execution, and the uncertainties from computational resources performances that may impose significant overhead delays.

WaaS cloud platforms extend the capabilities of a traditional Workflow Management System (MWS) to provide a more comprehensive service for larger users. In a traditional WMS, a single workflow job is processed in dedicated service to ensure its Quality of Service (QoS) requirements. However, this approach may not be able to cope up with the WaaS environments. A significant deficiency may arise from its conventional way of tailoring workflows' software configurations into a VM image, intra-dependent workflows' tasks inevitable schedule gaps, and possible overhead delays from workflow pre-processing and data movement handling.

To achieve more efficient multi-tenant WaaS cloud platforms, we proposed a novel resource-sharing policy that utilizes container technology to wrap software configurations required by particular workflows. This strategy enables (i) the idle slots VMs sharing between users and (ii) gaining further makespan minimization by sharing the datasets and container images cached within VMs local storage. We implemented this policy on EBPSM, a dynamic heuristic scheduling algorithms designed for multiple workflows scheduling in WaaS. Our extensive experiments show that the sharing scenarios overthrow a traditional dedicated approach in handling workflows' jobs. Furthermore, our proposed algorithm can surpass a modified state-of-the-art budget-constrained dynamic scheduling algorithm in terms of minimizing the makespan of workflows and meeting the soft limit budget defined by the users.

There are several aspects of our experiments that need to be further investigated. First, the budget distribution phase takes a vital role in budget-constrained scheduling. The decision to either allocate more budget for the earlier tasks so they can lease faster computational resources or maintain a safety net allocation to ensure the budget compliance must be carefully taken into account. A trade-off between having a faster execution time and meeting the allocated budget is always inevitable. In this way, defining the nature of execution, including strictness of the budget constraints can help to design an appropriate configuration between two approaches.

Second, the resource provisioning (and deprovisioning) strategy must consider the \textit{quid pro quo} between having a higher system utilization (i.e., lower idle VM times) and an optimal data sharing and movement which utilizes the VM local storage. We observed that delaying a particular VM termination in the deprovisioning phase may improve the performance when the cached data stored within the VM is intelligently considered. In this work, we did not consider task failure either caused by the software or the infrastructure (i.e., container, VMs). Incorporating a fault-tolerant strategy into both EPSM and EBPSM algorithms is necessary to address the task failure that highly likely impacting the WaaS cloud platforms performance. 

Finally, further investigation on how multiple container instances can be run and scheduled on top of a single VM is another to-do list. The delay in initiating a container image has reduced our algorithm's performance. There must be a way to counterbalance this issue by exploiting the swarming nature of container to gain benefits from this predetermined condition to enhance the WaaS cloud platforms efficiency further.

\section*{Acknowledgments}
	
	This research is partially supported by LPDP (Indonesia Endowment Fund for Education) and ARC (Australia Research Council) research grant.
	
	\bibliography{mybibfile}
	
\end{document}